\documentclass[aps,prb,twocolumn,showpacs,superscriptaddress,floatfix,10pt]{revtex4-1}
\usepackage{amsmath,amssymb,amsfonts,stmaryrd,wasysym,graphicx,multirow,color,textcomp,subfigure}
\usepackage{url}\usepackage{diagrams}
\usepackage[colorlinks=true,citecolor=blue,urlcolor=blue]{hyperref}
\usepackage{romannum}

\def\be{\begin{equation}}
\def\ee{\end{equation}}
\def\bea{\begin{eqnarray}}
\def\eea{\end{eqnarray}}
\def\d3{D\Romannum{3}}
\begin{document}
\pagenumbering{arabic}
\title{Coupled Wire Models of Interacting Dirac Nodal Superconductors}

\author{Moon Jip Park}
\affiliation{Department of Physics, University of Illinois at Urbana-Champaign, Urbana, IL 61801, USA}
\author{Syed Raza}
\affiliation{Department of Physics, University of Virginia, Charlottesville VA 22904, USA}
\author{Matthew J. Gilbert}
\affiliation{Department of Electrical and Computer Engineering, University of Illinois at Urbana-Champaign, Urbana, IL 61801, USA}
\affiliation{Department of Electrical Engineering, Stanford University, Stanford, CA 94305, USA}
\author{Jeffrey C. Y. Teo}
\affiliation{Department of Physics, University of Virginia, Charlottesville VA 22904, USA}
\date{\today}

\begin{abstract}
Topological nodal superconductors possess gapless low energy excitations that are characterized by point or line nodal Fermi surfaces. In this work, using a coupled wire construction, we study topological nodal superconductors that have protected Dirac nodal points. In this construction, the low-energy electronic degrees of freedom are confined in a three dimensional array of wires, which emerge as pairing vortices of a microscopic superconducting system. The vortex array harbors an antiferromagnetic time-reversal and a mirror glide symmetry that protect the massless Dirac fermion in the single-body non-interacting limit. Within this model, we demonstrate exact-solvable many-body interactions that preserve the underlying symmetries and introduce a finite excitation energy gap. These gapping interactions support fractionalization and generically lead to non-trivial topological order. We also construct a special case of $N=16$ Dirac fermions where corresponding the gapping interaction leads to a trivial $E_8$ topological order that is closely related to the cancellation of the large gravitational anomaly.
\end{abstract}

\maketitle

\section{Introduction}
Soon after the discovery of the topological band insulators\cite{RevModPhys.82.3045,RevModPhys.83.1057}, generalizing the topological phases to various materials has been one of the most popular themes in condensed matter physics\cite{RevModPhys.88.035005,doi:10.1002/pssr.201206451,doi:10.1146/annurev-conmatphys-031214-014749,doi:10.1146/annurev-conmatphys-031214-014501}. One intensively considered path of extending the topological phases is to consider the topological properties of semimetallic phases. Topological semimetallic phases possess a bulk degeneracy that is protected by the presence of an underlying topology. Up to now, the $3D$ topological semimetals are largely classified into the two classes: Weyl semimetals and Dirac semimetals. Weyl semimetals have two-fold linear band crossings and generally come in two interconnected varieties, namely type-1 and type-2. Type-1 Weyl semimetals have  either broken time-reversal or inversion symmetry and have been found in non-centrosymmetric materials such as: $\mathrm{TaAs}$\cite{PhysRevX.5.031013}, $\mathrm{TaP}$, $\mathrm{NbP}$, and $\mathrm{NbAs}$\cite{Liu2015,Xu2015}. Type-2 Weyl semimetals possess an additional broken Lorentz invariance and both $\mathrm{MoTe_2}$\cite{MoTe,MoTe2} and $\mathrm{WTe_2}$\cite{WTe,WTe2} are observed to be the type-2 Weyl semimetals\cite{type2}. Weyl semimetals have been predicted to have numerous distinguishing physical responses related to the presence of the chiral anomaly\cite{PhysRev.177.2426,Bell1969,Nielsen1983}. Examples of anomalous behavior in Weyl semimetals include: nonlocal quasiparticle transport\cite{PhysRevX.4.031035}, chiral magnetic effect\cite{Li2016,PhysRevB.92.161110,PhysRevB.89.081407}, chiral vortical effect\cite{PhysRevB.89.075124}, angular dependence of the magenetoresistence\cite{PhysRevX.5.031023,PhysRevB.88.104412}.

 Unlike the Weyl semimetals, the Dirac semimetals have four-fold degeneracy and require additional symmetries for the topological protection of the gapless bulk Dirac point. Most Dirac semimetals are found in non-magnetic materials such as $\mathrm{Na_3Bi}$\cite{Liu864,PhysRevB.85.195320,Liu864,Xu294,Xiong413} and $\mathrm{Cd_3 As_2}$\cite{Liu2014,PhysRevB.88.125427,Neupane2014,PhysRevLett.113.027603,Jeon2014,Liang2014,PhysRevLett.113.246402,PhysRevLett.115.226401,PhysRevB.92.081306,Li2015,Li2016,Guo2016,Zhang2017}, that preserve both time-reversal symmetry and inversion symmetry. Recently, the discovery of the Dirac semimetals has been extended to include the antiferromagnetic material $\mathrm{CuMnAs}$ that breaks both inversion and time-reversal symmetries yet preserves the product of the two\cite{Tang2016}. As is the case with Weyl semimetals, Dirac semimetals are predicted to possess physical manifestations that are separate and distinct from those found in Weyl semimetals or a $\mathbb{Z}_2$ anomaly\cite{PhysRevLett.117.136602,Xiong413}.

The study of the gapless topological phases can be further generalized into the class of superconducting states, often referred to as topological nodal superconductors\cite{PhysRevB.90.205136,1367-2630-15-6-065001,PhysRevLett.110.240404,0953-8984-27-24-243201}. The topological nodal superconductors are the superconducting analogue of the topological semimetals. The topological nodal superconductors possess nodal points or lines in the Brillouin zone(BZ), which has the vanishing superconducting gap. There has been numerous experimental and theoretical studies of the line nodal superconductors such as noncentrosymmetric superconductors including: $\mathrm{CePt_3Si}$\cite{PhysRevLett.94.207002,PhysRevLett.94.197002}, $\mathrm{Li_2Pt_3B}$\cite{PhysRevLett.97.017006}, and $\mathrm{CeIrSi_3}$\cite{PhysRevLett.100.107003}, and the heavy fermion compounds, $\mathrm{UBe_{13}}$\cite{PhysRevLett.52.1915}. Point nodal superconductors, often referred to as Weyl superconductors, are also proposed to exist in a veritable plethora of materials and systems including: $\mathrm{A}$ phase of $\mathrm{^{3}He}$\cite{RevModPhys.47.415,Volovik2011,volovik}, topological insulator-superconductor multilayers\cite{PhysRevB.86.054504}, doped Weyl semimetals\cite{PhysRevLett.120.067003,PhysRevB.86.214514,PhysRevB.92.035153,Huang2017,PhysRevB.93.184511}, the $\mathrm{B}$ phase of $\mathrm{UPt_3}$\cite{PhysRevB.92.214504}, the pnictide material $\mathrm{SrPtAs}$ \cite{PhysRevB.89.020509}, ferromagnetic superconductors\cite{PhysRevB.86.104509}, the superfluidity of Fermi gases\cite{PhysRevLett.115.265304,PhysRevLett.114.045302}, mirror symmetric superconductors\cite{PhysRevB.97.060504}, the half-metal/$d$-wave superconductor heterostructure\cite{PhysRevB.95.064513}, $\mathrm{Nb}$-doped $\mathrm{Bi_2Se_3}$\cite{PhysRevB.94.180510,PhysRevB.95.201109,PhysRevB.95.201110}, the $\mathrm{Cu}$-doped $\mathrm{Bi_2Se_3}$\cite{PhysRevB.96.144512,PhysRevB.94.180504,PhysRevLett.113.046401}, $\mathrm{PrOs_4Sb_{12}}$\cite{PhysRevLett.90.117001,PhysRevB.76.054514,ABUALRUB20081178}. $\mathrm{Cu_xBi_2Se_3}$ has been predicted to have the four-fold degenerate Dirac points, which is known as the Dirac superconductor\cite{PhysRevLett.113.046401}. As is evidenced in the above listed examples, topological nodal superconductors are often found in the strongly-correlated materials since the anisotropic pairing symmetries in the unconventional superconductors naturally introduces the nodal structures of the superconducting gap\cite{0953-8984-27-24-243201}. Therefore, it is crucial to study the strongly correlated phases of the nodal superconductors to fully understand the physical behavior of these materials.

In this regard, we study the superconductor analogue of the Dirac semimetals, namely Dirac nodal superconductors, in the presence of many-body interactions. To do so, we utilize the coupled wire construction method. In the coupled wire construction, the two and three dimensional phases of matter can be constructed by assembling an array of one dimensional wires. In this method, the many-body interactions are treated between neighboring wires, thereby it enables us to use the theoretical techniques that are only available in one dimension. This method has successfully reproduced and identified elementary excitations and behaviors of the numerous topological phases. In the two dimensional materials, the examples include the Laughlin states\cite{PhysRevLett.50.1395} and the hierarchy states\cite{PhysRevLett.51.605} of the fractional quantum Hall phases\cite{PhysRevLett.88.036401}, general Abelian and non-Abelian fractional quantum Hall phases\cite{PhysRevB.89.085101,PhysRevX.7.031009,Klinovaja2014,Meng2014}, fractional helical liquid{\cite{PhysRevB.89.115402}}, $2D$ fractional topological insulators\cite{PhysRevB.90.205101,PhysRevB.90.115426,PhysRevB.90.201102,PhysRevB.91.205141,PhysRevX.5.011011}, topological superconductors\cite{PhysRevX.4.011036,PhysRevB.89.104523} the surface of fractional topological insulator\cite{PhysRevB.90.201102,PhysRevX.5.011011}, and the spin-liquids\cite{PhysRevB.91.241106,PhysRevB.94.195130,PhysRevB.91.245139}.
The studies of the coupled wire construction even extend to the three-dimensional materials including $3D$ fractional topological phases\cite{PhysRevB.92.195137,PhysRevB.93.195136,Iadecola2017}, interacting Weyl semimetals\cite{PhysRevB.94.155136,0295-5075-102-6-67011,PhysRevB.92.115152,Mross}, the surface of $3D$ topological superconductor\cite{PhysRevB.94.165142}, and interacting Dirac semimetals\cite{Raza2017}.

\subsection{Summary of Results}\label{sec:introsummary}

In this work, we construct the many-body gapping potentials that generate a finite energy gap while preserving the underlying symmetries. In the presence of the many-body interactions, we find the emergence of the non-trivial topological orders.
We begin our discussion in Section \ref{sec:coupled} where we detail our construction the coupled wire model of a Dirac nodal superconductor in three spatial dimensions by assembling a vortex array in a microscopic superconductor within the continuum limit. In the continuum model, the massless Dirac fermions are protected by the combination of: local time-reversal symmetry, particle-hole symmetry and glide mirror symmetry. By introducing the array of superconducting pairing vortices, the low-energy electronic degrees of freedom manifest as $(1+1)$-D chiral Dirac fermions that are localized along vortex lines (also referred to as Dirac strings).  Each Dirac string is coupled via single-body tunneling with the adjacent strings, and the couplings reconstruct the Dirac nodal superconductor within the context of the coupled wire model. This anisotropic Dirac nodal superconducting model is protected by the same set of symmetries except time-reversal now becomes non-local and antiferromagnetic. This re-construction enables us to study many-body interactions in three dimensions using bosonization techniques.

With our introduction to the single body physics of the coupled wire methodology complete, in Section \ref{sec:manybody1} we introduce the many-body interactions that preserve all the underlying symmetries. The basic strategy that we follow in this work is based on the bi-partitioning the $SO(2N)$ Kac-Moody current consisting of $N$ chiral Dirac fermions along a vortex (see equation (31)). For even $N$, the symmetric gapping interaction can be facilitated by a simple separation $SO(2N)_1\sim SO(N)_1\times SO(N)_1$ of Dirac channels. The model admits a single-body mean-field mass gap, which reflects its trivial topology under the $\mathbb{Z}_2$  classification. On the other hand, due to the presence of the aforementioned symmetries, the odd $N$ case requires a non-trivial string decomposition that involves the level-rank duality $SO(9)_1\sim SO(3)_3\times SO(3)_3$. Consequently, the gapping interactions in the case of odd $N$ lead to fractionalization and non-trivial topological order. In both situations, the gapping potentials are constructed by backscattering the divided Kac-Moody currents to opposite directions between adjacent strings. This results in a finite energy gap while preserving all the underlying symmetries of our model.

Interestingly, when $N=16$, we find a special form of the decomposition, $SO(32)\sim E_8 \times E_8$, that utilizes the $E_8$ unimodular lattice. We find that this decomposition results in the many-body interaction that has trivial topological order. In an effort to understand this result more clearly, in Section \ref{secmodreview}, we utilize modular transformations confirm the presence of a topologically trivial phase results from this decomposition is reflected by the cancellation of the large gravitational anomaly. Thereby, the absence of the gravitational anomaly signals the topologically trivial many-body gapping potential\cite{PhysRevB.85.245132,RevModPhys.88.035001,Ryu2013,Hsieh2014,Cappelli2013,Shinsei2015,PhysRevB.95.235130}.
\section{Coupled Wire Models of Dirac Nodal Superconductor}\label{sec:coupled}
In this section, we describe the single-body aspects of the Dirac nodal superconductor in terms of a coupled wire model. This mirrors the discussions of the non-interacting model in ref.~\onlinecite{RazaSirotaTeo17}. The major difference between our work and previous implementations of the coupled wire constructions of topological phenomena is that in this work we focus on superconducting media that break charge $U(1)$ conservation. The basic building blocks of the coupled wire model are chiral Dirac wires. These are $(1+1)$-D Dirac fermion channels where quasiparticles can only propagate in a single direction. They can be supported by an array of vortices in a degenerate point nodal superconductor in the continuum model. In our model, the nodal superconducting state has a normal metallic parent state that is quasi-one-dimensional with dispersion predominantly in the $y$-direction and pairing order that is $p$-wave directed along the normal $xz$-plane. The Hamiltonian of the nodal superconductor in the continuum limit is given as,
\begin{align}
H_{\mathrm{SC-nodal}}({\bf k})&=(\hbar vk_ys_y\mu_z-\epsilon_f)\tau_z\nonumber\\&\;\;\;\;+\Delta l(\tilde{k}_xs_z-\tilde{k}_zs_x)\mu_x\tau_x,
\label{DiracNODAL}
\end{align}
and is acting on the Nambu vector $\boldsymbol\xi=({\bf c},is_y{\bf c}^\dagger)^T$, where ${\bf c}=(c_{s\mu})$ are the (complex) Dirac fermion annihilation operators, and $s=\uparrow,\downarrow$ and $\mu=\pm$ are the Pauli matrices for the spin and Weyl-species degrees of freedom respectively. In addition, $\vec\tau=(\tau_x,\tau_y,\tau_z)$ are Pauli matrices acting on the Nambu $(c,c^\dagger)$ degree of freedom.

The model in Eq. \eqref{DiracNODAL} can be further simplified by the unitary basis transformation, $U=(\tau_z+s_y\mu_y\tau_x)/\sqrt{2}$ when the momenta are rescaled so that $\Delta l\tilde{k}=\hbar vk$, and the Fermi energy is set at $\epsilon_f=0$. Under these aforementioned conditions, we define a new, simplified superconducting Hamiltonian $H_{\mathrm{SC-Dirac}}=UH_{\mathrm{SC-nodal}}U^{-1}$, written explicitly as,
\begin{align}
H_{\mathrm{SC-Dirac}}({\bf k})=\hbar v{\bf k}\cdot\vec{s}\mu_z\tau_z,
\label{DiracSC}
\end{align}
where $\vec{s}=(s_x,s_y,s_z)$ are Pauli spin matrices and $\mu_z=\pm1$ indexes the two Weyl species with opposite Chern numbers.

The BdG Hamiltonian in Eq. \eqref{DiracSC} has the time-reversal symmetry,
\begin{align}
s_yH_{\mathrm{SC-Dirac}}({\bf k})^\ast s_y=H_{\mathrm{SC-Dirac}}(-{\bf k}),
\label{DiracNODALHTR}
\end{align} and the particle-hole symmetry $s_y\tau_yH({\bf k})^\ast s_y\tau_y=-H(-{\bf k})$ due to the Nambu doubling. In addition, we consider the effective mirror glide symmetry,
\begin{align}
s_z\tau_yH_{\mathrm{SC-Dirac}}({\bf k})s_z\tau_y=H_{\mathrm{SC-Dirac}}(M{\bf k}),
\label{DiracNODALHglide}
\end{align}
where $M:(k_x,k_y,k_z)\to(k_x,k_y,-k_z)$. The mirror-glide $G=s_z\tau_y$ is consistent with the Nambu doubling as it commutes with the particle-hole operator $\Xi=s_y\tau_y\mathcal{K}$, where $\mathcal{K}$ is the complex conjugation operator. The mirror glide $G$ and time-reversal $T=is_y\mathcal{K}$ also mutually commute. While a symmorphic mirror operator squares to $-1$ in a spinful system, a nonsymmorphic glide operator squares to $-e^{iK\cdot{\bf a}}$, where ${\bf a}$ is a microscopic in-plane lattice translation. In our model, we assume the Dirac degeneracy sits at the microscopic lattice momentum $K$ so that $e^{iK\cdot{\bf a}}=-1$, and the Hamiltonian \eqref{DiracNODAL} describes the small momentum deviation ${\bf k}$ away from $K$ in the long length-scale limit ${\bf a}\to0$. The time-reversal and particle-hole operator are unaltered by the basis transformation, $U$, but the glide symmetry changes from $G=-s_x\mu_y$ to $G=s_z\tau_y$. For mathematical convenience, we will utilize the non-electronic basis where the BdG Hamiltonian is Eq. \eqref{DiracSC} and $G=s_z\tau_y$.

The time-reversal and particle-hole symmetry allow the gap-opening mass terms $\Delta\mu_z\tau_x+m_1\tau_x+m_2\mu_x\tau_z$ but neither of these terms preserves the glide symmetry. We notice in passing that these glide breaking mass terms can lead to a set of interesting topological superconducting states in the Altland-Zirnbauer DIII class~\cite{AltlandZirnbauer97}. A non-vanishing energy gap arises when $|\Delta|\neq\sqrt{m_1^2+m_2^2}$ and there are three disconnected regions separated by the condition where $|\Delta|=\sqrt{m_1^2+m_2^2}$. The two disconnected regions defined by $\Delta>\sqrt{m_1^2+m_2^2}$ and $\Delta<-\sqrt{m_1^2+m_2^2}$ are occupied by time-reversal symmetric topological superconductors~\cite{Ryu2008,Kitaevtable08} with topological indices $N=1$ and $-1$ respectively. The remaining region $|\Delta|<\sqrt{m_1^2+m_2^2}$ is path-connected and is occupied by trivial superconductors with topological index $N=0$. However, the region $|\Delta|<\sqrt{m_1^2+m_2^2}$ is not simply-connected. There is a fundamental homotopy group $\pi_1=\mathbb{Z}$, which classifies vortices of $m({\bf r})=m_1({\bf r})+im_2({\bf r})=|m|e^{i\varphi({\bf r})}$ where the phase $\varphi({\bf r})$ spatially modulates and winds $2\pi n$ around a vortex. A vortex line hosts $n$ pairs of helical Majorana fermions modes, which are protected by time-reversal symmetry when $n$ is odd.

\begin{figure}[htbp]
\centering\includegraphics[width=0.45\textwidth]{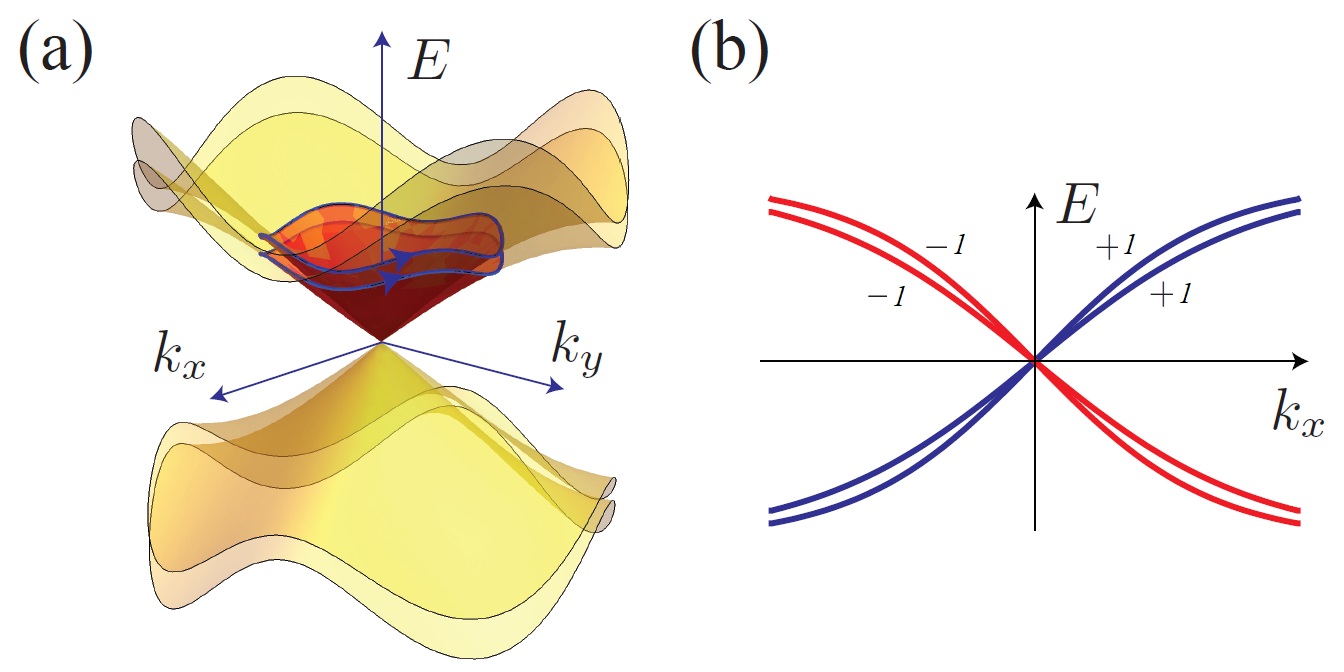}
\caption{(a) The energy spectrum of Dirac nodal superconductor along the mirror symmetric plane($k_z=0$). The Dirac nodal superconductor is protected by non-trivial mirror winding (blue curve) around the nodal point. (b) The same dispersion as (a) but now labeled with the mirror-glide eigenvalue(red and blue), $g$. Each mirror branch now connects at higher momentum, and they switch each other.}\label{fig:MajoranaCone}
\end{figure}

We now restrict our model to be the mirror glide symmetric. The nodal superconducting Dirac state in Eq. \eqref{DiracSC} is stable in the single-body setting and is protected by time-reversal symmetry, $T$, and mirror-glide symmetry, $G$. This can be verified by explicitly checking that there are no symmetry-preserving gap-opening mass terms. Alternatively, this can also be explained using topological reasoning that does not require the specific form of the Hamiltonian. Let us begin by focusing on the mirror-glide symmetric $k_x-k_y$ plane where $k_z=0$. Along this plane, the BdG Hamiltonian commutes with the mirror-glide operator $G=s_z\tau_y$ and, thus, can be block diagonalized according to the corresponding mirror-glide eigenvalues $g=\pm1$. Since $G$ commutes with both time-reversal symmetry and particle-hole symmetry, each mirror sector also carries the same symmetries.  Each sector consists of a protected pair of massless Majorana fermions in two dimensions -- equivalent to those living on the surface of a class DIII topological superconductor~\cite{Ryu2008,Kitaevtable08} with topological index $N=\pm2$(See Figure~\ref{fig:MajoranaCone} (a))., where the sign depends on the mirror-glide eigenvalue, $g$. Unlike the topological surface state which is anomalous, the nodal superconducting state here does not require a higher dimensional bulk. This is because of the following two reasons: First, the winding numbers of the two mirror-glide sectors are opposite to one another and cancel. Second, the mirror-glide symmetry, $G$, is nonsymmorphic and as such squares to the translation phase $e^{i{\bf k}\cdot{\bf a}}$, where ${\bf a}$ is the microscopic in-plane lattice vector that has been taken to zero as a continuum limit. Consequently, the spectrum of $G$ is $\pm e^{i{\bf k}\cdot{\bf a}/2}$ instead of $\pm1$. The two eigenvalue branches connect and switch in momentum space across the microscopic Brillouin zone when ${\bf k}\to{\bf k}+{\bf G}$, where ${\bf G}$ is the reciprocal lattice vector dual to ${\bf a}$ (i.e.~${\bf G}\cdot{\bf a}=2\pi$). As a result, the two mirror-glide sectors are not decoupled as they also connect and switch at large momentum and small length-scale (See Figure~\ref{fig:MajoranaCone} (b)). We notice the resemblance to the charge conserving Dirac (semi)metal, which is protected by time-reversal and screw rotation symmetries~\cite{RazaSirotaTeo17}.  We refer to the current case as a Dirac nodal superconductor in three dimensions.

As the nodal points in the $3$D Dirac topological superconductor are topologically protected, we now define and evaluate a topological mirror-glide winding invariant. Since both the time-reversal and particle-hole operators commute with the mirror-glide matrix $G$, so does the chiral operator, which is the product $S=-iT\Xi=\tau_y$. Let ${\bf v}_{sg}^1$ and ${\bf v}_{sg}^2$ be two orthonormal simultaneous eigenvectors of $S$ and $G$ with eigenvalues $s=\pm1$ and $g=\pm1$ respectively. Then, we can define the projection operator, $P_{sg}^\dagger=({\bf v}_{sg}^1,{\bf v}_{sg}^2)^\dagger$, that maps onto the two-dimensional fixed eigenvalue subspace. Since the BdG Hamiltonian $H_{\mathrm{SC-Dirac}}$ in Eq. \eqref{DiracSC} commutes with $G$ on the $k_x-k_y$ plane and anticommutes with $S$, it can be summarized by the two $2\times2$ matrices \begin{align}h^{(+)}({\bf k}_\|)&=P_{-+}^\dagger H_{\mathrm{SC-Dirac}}({\bf k}_\|)P_{++},\nonumber\\h^{(-)}({\bf k}_\|)&=P_{--}^\dagger H_{\mathrm{SC-Dirac}}({\bf k}_\|)P_{+-}\end{align} associated with the two distinct mirror sectors, where ${\bf k}_\|$ is the in-plane momentum. As the Hamiltonian is nodal only at the $\Gamma$ point, $h^{(\pm)}({\bf k}_\|)$ is non-singular as long as ${\bf k}_\|\neq0$. Therefore, the winding number of each sector is defined to be the integral \begin{align}N^{(\pm)}=\frac{1}{2\pi i}\oint_{\mathcal{C}}\mathrm{Tr}\left[h^{(\pm)}({\bf k}_\|)^{-1}\nabla_{{\bf k}_\|}h^{(\pm)}({\bf k}_\|)\right]\cdot d{\bf k}_\|\label{Mwinding}\end{align} where $\mathcal{C}$ can be taken to be any (anti-clockwise) loop around the origin on the momentum plane. This integral represents the same winding invariant that characterizes the surface Majorana cone of a time-reversal invariant topological superconductor. This analysis confirms that along the mirror symmetric plane, the gapless Majorana fermions corresponding to each mirror-glide sector are equivalent to those on the surface of a time-reversal invariant topological superconductor with index $N=\pm2$\cite{Ryu2008}.  For the Dirac nodal superconductor in Eq. \eqref{DiracSC}, these winding numbers are $N^{(+)}=-N^{(-)}=2$. The two sectors must have opposite invariants as the overall system is anomalous-free. In general, we define the mirror winding number to be $N=N^{(+)}=-N^{(-)}$. Since the winding number in Eq. \eqref{Mwinding} must be an integer and cannot change continuously, a nodal superconductor with non-trivial mirror winding is topological protected as long as the symmetries are preserved. It is important to note that the glide mirror symmetry in our model squares to $1$, which is different from the conventional mirror symmetry in a spinful system. If $G^2=-1$ (up to a translation), the eigenvalues of $G$ are $\pm i$. In this case, we can define the mirror wining number of $G=+i$ and $G=-i$ sectors respectively. However $T$ flips between the two sectors of $G$. Therefore the winding number of the $G=+i$ branch must be equal to that of the $G=-i$ branch. So the net winding number is now non-zero. As a result, the Dirac nodal superconductor must be anomalous, and as such it can only appear at the boundary of a 4D bulk.

We now consider the presence of chiral Dirac vortices in the 3D Dirac nodal superconductor that break both the time-reversal and mirror-glide symmetries. Each of these vortices host a single chiral (complex) Dirac fermion. They constitute a three dimensional array of coupled vortices that restores the symmetries in low-energy long length-scale. The BdG defect Hamiltonian of the vortex array consists of the nodal Hamiltonian in Eq. \eqref{DiracSC} together with the symmetry breaking terms
\begin{align}
H_{\mathrm{SC-Dirac}}({\bf k},{\bf r})&=\hbar v{\bf k}\cdot\vec\sigma\mu_z\tau_z\nonumber\\&\;\;\;\;+\Delta_1({\bf r})\tau_x+\Delta_2({\bf r})\mu_z\tau_y.
\label{varray}
\end{align} $\Delta({\bf r})=\Delta_1({\bf r})+i\Delta_2({\bf r})$ is the order parameter that acts independently on the two Weyl species labeled by $\mu_z=\pm1$ and slowly modulates in space. When $\Delta({\bf r})=|\Delta|e^{i\varphi({\bf r})}$ forms a vortex configuration in that its phase $\varphi$ spatially winds by $2\pi$ around the defect line. Each Weyl sector in the Hamiltonian supports a chiral (real) Majorana channel along the vortex lines. These are quasi-one-dimensional structures that host gapless Majorana fermion excitations. The Majorana fermions are localized on the vortex line and they are chiral in the sense that they can only propagate along a single direction. The superconducting pairing potentials in Eq. \eqref{varray} are chosen so that the phases of the order parameter are conjugated between the two Weyl species. Together with the fact that the two Weyl species have the opposite Chern numbers, the pair of chiral Majorana channels supported by them are propagating in the same direction and are protected. For instance, electron tunnelings between the pair are forward scattering processes that only renormalize the velocity and cannot introduce a mass. The pair of co-propagating Majorana fermions can be combined into a single chiral Dirac fermion $d\sim\gamma_++i\gamma_-$.

\begin{figure}[htbp]
\centering\includegraphics[width=0.3\textwidth]{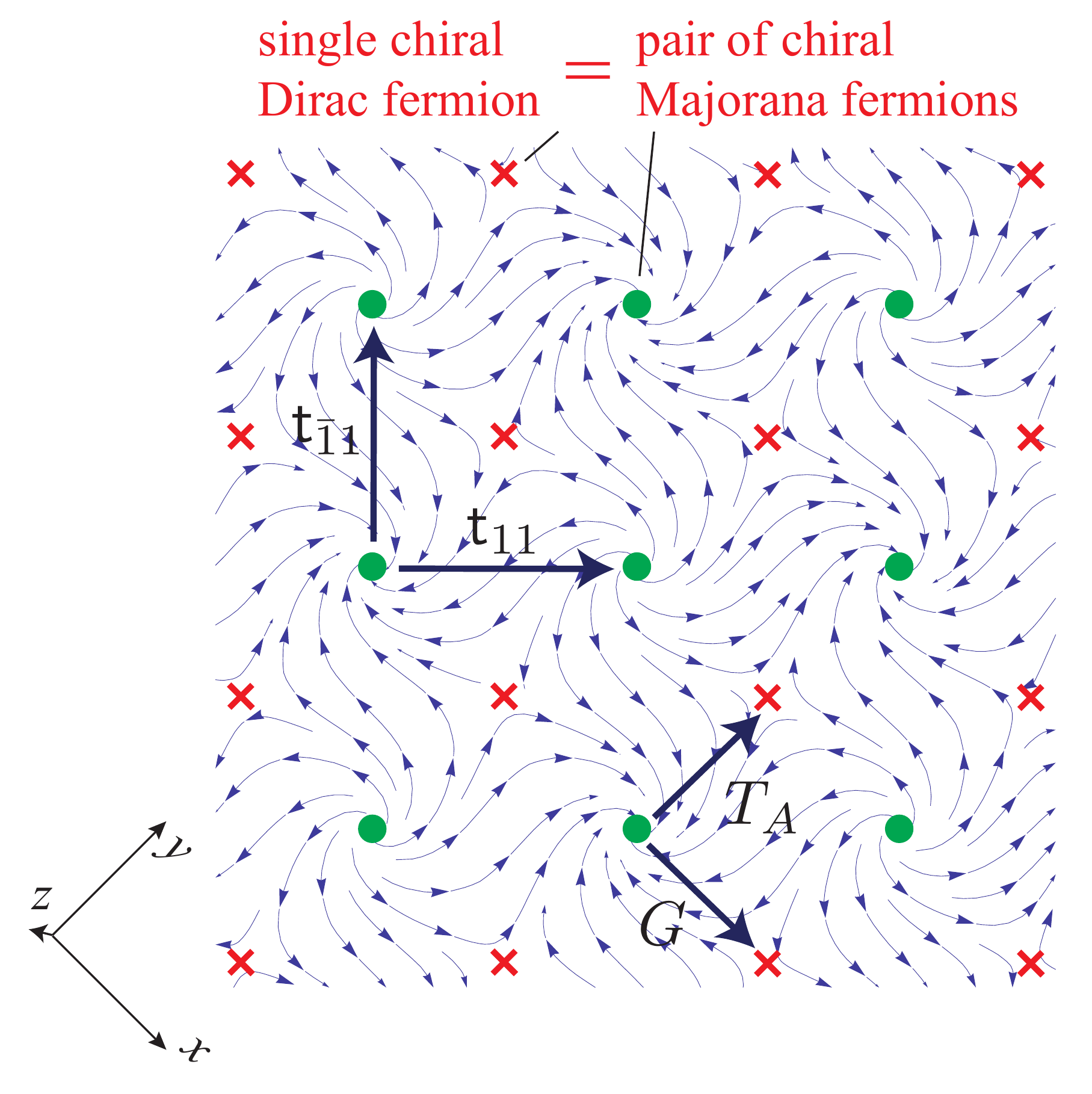}
\caption{Arrays of chiral Dirac channels that we use as a basis for constructing the 3D nodal Dirac superconductor in this work. The vortex array is symmetric under antiferromagnetic time-reversal $T_A$ and mirror-glide $G$. The vector field represents the pairing phase of the host superconducting state. Each {\color{green}$\bullet$} and {\color{red}$\times$} represents a neutral Dirac fermion channel parallel and opposite to the $z$-axis.}\label{fig:chiralDiracvortices}
\end{figure}

A periodic array of chiral Dirac vortices can be realized by the pairing configuration
 \begin{align}
\Delta({\bf r})=\Delta_0\frac{\mathrm{sd}(x+iy)}{|\mathrm{sd}(x+iy)|}
\end{align} where $\mathrm{sd}$ is the (rescaled) Jacobian elliptic function~\cite{ReinhardtWalker10} with simple zeros and poles at $p+iq$ and $(p+1/2)+i(q+1/2)$ respectively, $p,q$ are integers. Fig. \ref{fig:chiralDiracvortices} shows the checker board lattice, and ${\bf e}_x$ and ${\bf e}_y$ form a lattice translational vector. We find that the phase $\varphi$ of $\Delta=|\Delta|e^{i\varphi}$ winds by $2\pi$ around each integral lattice point and $-2\pi$ around a half-integral one. The vortex lines are directed along $z$ and form a checkerboard lattice along $x$ and $y$. Nearest neighbor vortices host counter-propagating Dirac modes so that a right-moving chiral Dirac channel appears at $(p,q)$ and a left-moving one appears at $(p+1/2,q+1/2)$.
 We choose $\Delta_0=|\Delta_0|e^{i\pi/4}$ so that the pairing order parameter transforms under
 \begin{gather}
 \Delta({\bf r})=-\Delta({\bf r}+{\bf e}_x)=-\Delta({\bf r}+{\bf e}_y),\nonumber\\\Delta\left({\bf r}+\frac{{\bf e}_x\pm{\bf e}_y}{2}\right)=\pm\Delta({\bf r})^\ast.
 \end{gather}
 Although the pairing order parameter has periods $(1,1)$ and $(1,-1)$, the Bogoliubov-de Gennes (BdG) Hamiltonian in Eq. \eqref{varray} has finer artificial lattice translation symmetries $\mathsf{t}_x,\mathsf{t}_y$. The Hamiltonian symmetric under $\mathsf{t}_x,\mathsf{t}_y$ satisfies,
 \begin{align}
 \mu_y\tau_yH_{\mathrm{SC-Dirac}}({\bf k},{\bf r})\mu_y\tau_y&=H_{\mathrm{SC-Dirac}}({\bf k},{\bf r}+{\bf e}_x)\nonumber\\&=H_{\mathrm{SC-Dirac}}({\bf k},{\bf r}+{\bf e}_y).
 \label{latticetranslationH}
 \end{align}
 The Hamiltonian is symmetric under particle-hole operator, $\Xi$, and it follows the condition given as,
 \begin{align}s_y\tau_yH_{\mathrm{SC-Dirac}}({\bf k},{\bf r})^\ast s_y\tau_y=-H_{\mathrm{SC-Dirac}}(-{\bf k},{\bf r}).\end{align}
 Furthermore, our particular vortex geometry possesses an antiferromagnetic time-reversal $T_A$, which is a combination of the time reversal $T$ and a half-translation by $({\bf e}_x+{\bf e}_y)/2$.
 \begin{align}s_yH_{\mathrm{SC-Dirac}}({\bf k},{\bf r})^\ast s_y=H_{\mathrm{SC-Dirac}}\left(-{\bf k},{\bf r}+\frac{{\bf e}_x+{\bf e}_y}{2}\right).\label{TAH}
 \end{align}
 The vortex array also possesses the mirror-glide symmetry $G$, which combines mirror along the $x-y$ plane with a half-translation by $({\bf e}_x-{\bf e}_y)/2$,
 \begin{align}&s_z\tau_yH_{\mathrm{SC-Dirac}}({\bf k},{\bf r})s_z\tau_y\nonumber\\&\;\;\;=H_{\mathrm{SC-Dirac}}\left(M{\bf k},M{\bf r}+\frac{{\bf e}_x-{\bf e}_y}{2}\right),\label{GH}\end{align}
 where we define the mirror symmetry operator as, $M:(x,y,z)\to(x,y,-z)$.

We now focus on the low-energy chiral Dirac modes in the vortex array. Let $d_{x,y,k}$ be the chiral Dirac mode at $(x,y)$ with momentum $k_z=k$. When $x\equiv y$ modulo 2, $d_{x,y,k}=R_{x,y,k}$ propagates in the $+z$ direction, and when $x\equiv y+1$ modulo 2, $d_{x,y,k}=L_{x,y,k}$ propagates in the $-z$ direction. $R$ and $L$ are respectively represented by crosses and dots in Figure ~\ref{fig:chiralDiracvortices}. Each is a combination of a pair of chiral Majorana modes that are originated from the two opposite Weyl species. $R\sim\gamma_R^{(+)}+i\gamma_R^{(-)}$, where the sign refers to chirality of the original Weyl species $\mu_z=\pm1$, and similar definitions extend to the $-z$ directed modes as $L\sim\gamma_L^{(+)}+i\gamma_L^{(-)}$. Due to the presence of the underlying symmetries in the system, the Dirac fermions transform according to \begin{gather}\mathsf{t}_{11}R_{x,y,k}\mathsf{t}_{11}^{-1}=R_{x+1,y+1,k},\quad\mathsf{t}_{\bar{1}1}R_{x,y,k}\mathsf{t}_{\bar{1}1}^{-1}=R_{x-1,y+1,k},\nonumber\\\mathsf{t}_{11}L_{x,y,k}\mathsf{t}_{11}^{-1}=L_{x+1,y+1,k},\quad\mathsf{t}_{\bar{1}1}L_{x,y,k}\mathsf{t}_{\bar{1}1}^{-1}=L_{x-1,y+1,k},\nonumber\\T_AR_{x,y,k}T_A^{-1}=L_{x,y+1,-k},\quad T_AL_{x,y,k}T_A^{-1}=-R_{x,y+1,-k},\nonumber\\GR_{x,y,k}G^{-1}=iL_{x+1,y,-k}^\dagger,\quad GL_{x,y,k}G^{-1}=iR_{x+1,y,-k}^\dagger,\label{CWsymmetries}\end{gather} They form a representation that is consistent with the symmetry algebra $[\mathcal{O},\mathcal{O}']=0$ for $\mathcal{O},\mathcal{O}'=T_A,G,\mathsf{t}_{11},\mathsf{t}_{\bar{1}1}$, $T_A^2=(-1)^F\mathsf{t}_{11}\mathsf{t}_{\bar{1}1}$ and $G^2=\mathsf{t}_{11}\mathsf{t}_{\bar{1}1}^{-1}$, where $(-1)^F$ is the fermion parity operator and $T_A$ is antiunitary.(See Appendix \ref{Appendix:A} for the explicit form of the Dirac fermion operators).

\begin{figure}[htbp]
\centering\includegraphics[width=0.48\textwidth]{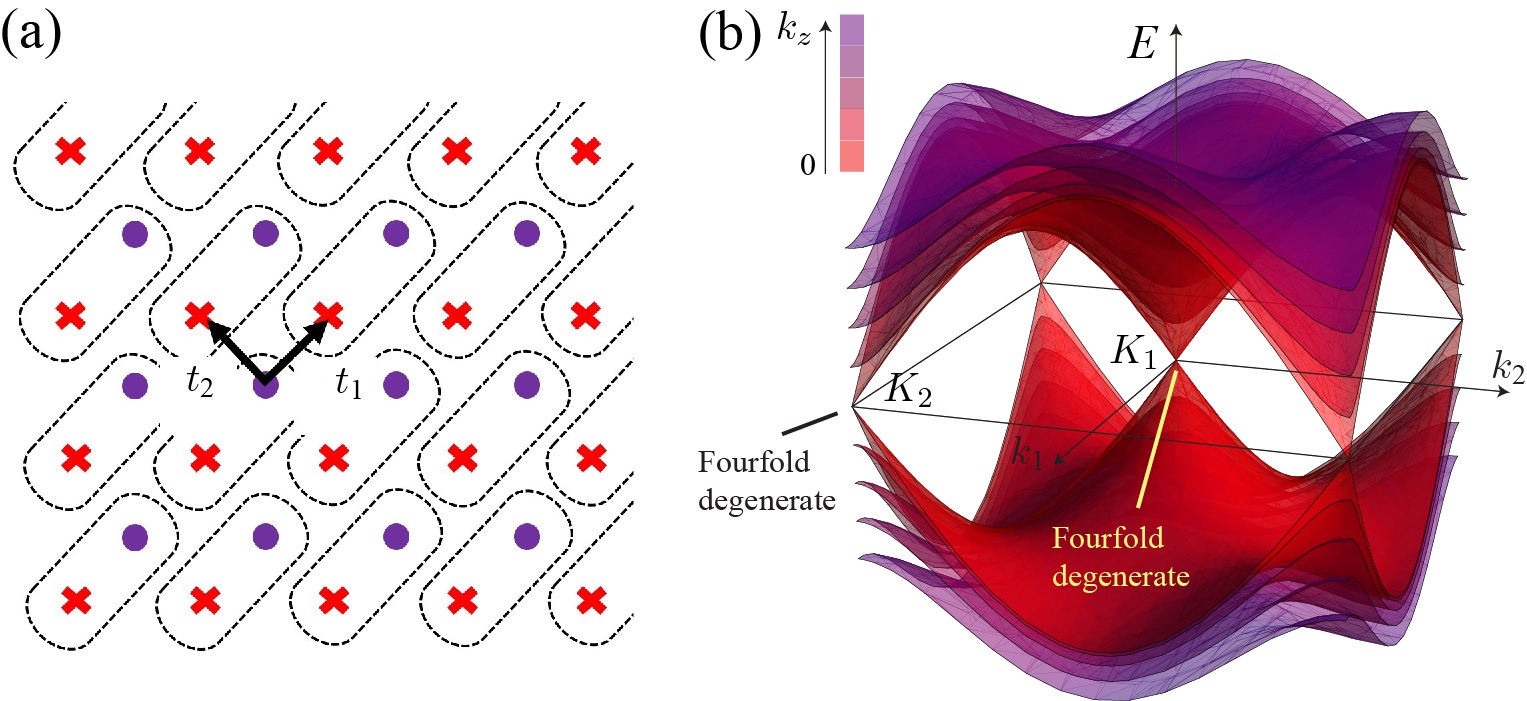}
\caption{(a) Schematic figure of the coupled wire model that we use in this work. Each unit cell (as denoted by the dashed black boxes) consists of a pair of counter-propagating Dirac modes, labeled by red cross and purple dot. The black arrows indicate tunneling amplitudes $t_1$ and $t_2$ between different Dirac modes. (b) Energy spectrum of the BdG model \eqref{Eq:BdG} of the coupled wire model. We find the two nodal points, $K_1$ and $K_2$, showing the massless Dirac dispersions.}\label{fig:3Darray}
\end{figure}

Integrating out the gapped bulk modes far away from the Fermi energy, the low-energy effective Hamiltonian can be written as \begin{gather}H_{||}(k)=\hbar v_fk\sum_{x\equiv y}\left(R_{x,y,k}^\dagger R_{x,y,k}-L_{x+1,y,k}^\dagger L_{x+1,y,k}\right)\label{Hparallel}\end{gather} where $v_f$ is the Fermi velocity of the Dirac fermions. When the wires are close to each other, there are finite hybridizations between the Dirac modes. We add interwire fermion quasiparticle tunneling terms in the $[110]$ and $[1\bar{1}0]$ directions. The presence of these terms in our model introduces an energy dispersion in the perpendicular directions. The symmetry-preserving interwire hopping terms can be written as,
\begin{gather}
H_{\perp,1}(k)=\sum_{x\equiv y}it_1\left(L^\dagger_{x,y+1,k}-L^\dagger_{x,y-1,k}\right)R_{x,y,k}+h.c.,
\\
H_{\perp,2}(k)=\sum_{x\equiv y}t_2\left(L^\dagger_{x-1,y,k}-L^\dagger_{x+1,y,k}\right)R_{x,y,k}+h.c.
\end{gather}
where the tunneling strengths $t_1$ and $t_2$ are real numbers. In Figure~\ref{fig:3Darray}(a), we schematically represent the physical picture of the counter-propagating modes with the corresponding interwire hopping terms, $t_1$ and $t_2$ that account for the interwire tunneling resulting from the spatial proximity between Dirac wires. By collecting the interwire hopping terms in both directions, the Hamiltonian of the coupled wire model is
\begin{align}\mathcal{H}&=\sum_{k}H_{||}(k)+H_{\perp,1}(k)+H_{\perp,2}(k).\label{Eq:WeylHam}\end{align} The model is symmetric under lattice translations, antiferromagnetic time-reversal, and mirror-glide symmetries as: \begin{align}\mathsf{t}_{11,\bar{1}1}^{-1}\mathcal{H}\mathsf{t}_{11,\bar{1}1}=T_A^{-1}\mathcal{H}T_A=G^{-1}\mathcal{H}G=\mathcal{H}.\end{align} After Fourier transform, the single-body system can be captured by the BdG Hamiltonian in the momentum space, \begin{align}\mathcal{H}&=\frac{1}{2}\sum_{k_1,k_2,k_z}\boldsymbol\xi_{\bf k}^\dagger H_{BdG}({\bf k})\boldsymbol\xi_{\bf k},\nonumber\\
H_{BdG}({\bf k})&=\begin{pmatrix}H({\bf k})&0\\0&-\sigma_yH(-{\bf k})^T\sigma_y\end{pmatrix},\label{Eq:BdG}\\
H({\bf k})&=\begin{pmatrix}
		  \hbar v_fk_z & q(k_1,k_2) \\
      q(k_1,k_2)^\ast & -\hbar v_fk_z \\
   \end{pmatrix},\nonumber
\end{align}
where $\boldsymbol\xi_{\bf k}=(R_{\bf k},L_{\bf k},L_{-{\bf k}}^\dagger,-R_{-{\bf k}}^\dagger)^T$ is the Nambu vector, $H({\bf k})$ is the Hamiltonian defined in Eq.~\eqref{Eq:WeylHam}, and $q(k_1,k_2)=it_1(1-e^{-i(k_1+k_2)})+t_2(e^{-ik_1}-e^{-ik_2})$. Here, $k_1$ is directed along the $(11)$-direction and $k_2$ is along the $(\bar{1}1)$-direction. From this point forward, we set $\hbar v_f=t_1=t_2=1$ (in units of $eV$) for simplicity but without loss of generality. In analyzing our model, we find that $q(0,0)=q(\pi,\pi)$ vanishes, therefore the Hamiltonian in Eq.~\eqref{Eq:WeylHam} possesses the two Weyl nodes at $K_1=(0,0,0)$ and $K_2=(\pi,\pi,0)$. We may write the low-energy expansion of the Hamiltonian near each of the Weyl points as,
\begin{gather}
H(K_{1,2}+{\bf k})\approx \nonumber
\\ \left(\begin{matrix}
      k_z & -t_1(k_1+k_2) \mp it_2(k_1-k_2) \\
      -t_1(k_1+k_2) \pm it_2(k_1-k_2) & -k_z \\
   \end{matrix}\right)\label{Eq:weylexpand}
\end{gather}
for small $|k| \ll 1$. From Eq.~\eqref{Eq:weylexpand}, we find that the Hamiltonian is comprised of a pair of Weyl fermions with opposite chiralities and we plot the resulting energy spectra in the low-energy limit in Figure \ref{fig:3Darray}(b).

Unlike the Weyl semimetals that has the charge conservation, our model is based on a charge breaking superconducting medium. The Weyl fermions here are not protected by the Chern number. This is because the BdG Hamiltonian in Eq. \eqref{Eq:BdG} has two opposing diagonal blocks. At each of the corresponding gap closing points $K_{1,2}$, there are two coinciding massless nodes in the BdG description and they have opposite Chern numbers. In fact, as the nodal points are inversion symmetric, where $K_{1,2}=-K_{1,2}$ (modulo reciprocal vectors), the particle-hole symmetry forbids a non-vanishing net Chern number. As a result, the Weyl fermions, in the absence of the symmetries, can acquire finite masses by the addition of off-diagonal terms in the BdG Hamiltonian of Eq. \eqref{Eq:BdG}. However, these terms are absent in our model because of the particle-hole, antiferromagnetic, and mirror-glide symmetries: \begin{align}\Xi H_{BdG}({\bf k})^\ast&=-H_{BdG}(-{\bf k})\Xi,\nonumber\\T_A({\bf k})H_{BdG}({\bf k})^\ast&=H_{BdG}(-{\bf k})T_A({\bf k}),\\G({\bf k})H_{BdG}({\bf k})&=H_{BdG}(M{\bf k})G({\bf k}),\nonumber\end{align} where the Nambu vector transforms according to \begin{align}\boldsymbol\xi_{\bf k}^\dagger&=\Xi\boldsymbol\xi_{-{\bf k}},\nonumber\\T_A\boldsymbol\xi_{\bf k}T_A^{-1}&=T_A({\bf k})\boldsymbol\xi_{-{\bf k}},\\G\boldsymbol\xi_{\bf k}G^{-1}&=G({\bf k})\boldsymbol\xi_{M{\bf k}},\nonumber\end{align} where $M$ is defined as $M:(k_1,k_2,k_z)\to(k_1,k_2,-k_z)$, and the symmetry matrices are given by \begin{gather}\Xi=\sigma_y\tau_y\nonumber\\T_A({\bf k})=\left(\begin{smallmatrix}0&1&0&0\\-e^{-i(k_1+k_2)}&0&0&0&\\0&0&0&e^{-i(k_1+k_2)}\\0&0&-1&0\end{smallmatrix}\right)\\G({\bf k})=\left(\begin{smallmatrix}0&0&ie^{-ik_2}&0\\0&0&0&-ie^{ik_1}\\-ie^{ik_1}&0&0&0\\0&ie^{-ik_2}&0&0\end{smallmatrix}\right).\nonumber\end{gather} The gapless nodes at $K_{1,2}$ are protected by the non-trivial mirror winding number $N(K_{1,2})=N^{(+)}(K_{1,2})=1$ defined in Eq. \eqref{Mwinding}. The winding numbers are equal at the two nodal momenta and, therefore, they add up to the net non-trivial mirror winding number $N=N(K_1)+N(K_2)=2$, which is identical to that of the homogeneous superconducting Dirac parent state in Eq. \eqref{DiracSC}. In other words, the coupled wire model recovers the Dirac nodal superconductor in low-energy limit. \begin{align}\begin{diagram}\stackrel{\mbox{Dirac nodal}}{\mbox{superconductor}}&\pile{\rTo^{\mbox{\small chiral vortices}}\\\lTo_{\mbox{\small coupled wire model}}}&\mbox{chiral Dirac strings}\end{diagram}\end{align}

We conclude this section by commenting the stability of Dirac nodal superconductors in the single-body setting. We first notice that the continuum model Hamiltonian, $H_{\mathrm{SC-Dirac}}$ in Eq. \eqref{DiracSC}, can be broken down into two pieces according to the Weyl species $\mu_z$. Each piece corresponds to a {\em Weyl nodal superconductor} \begin{align}H_{\mathrm{SC-Dirac}}=\pm\hbar v{\bf k}\cdot\vec{s}\tau_z,\label{WeylSC}\end{align} which is protected by time-reversal and glide symmetries and has the non-trivial mirror winding number $N=N^{(+)}=-N^{(-)}=1$. We are referring to Eq. \eqref{WeylSC} as a ``Weyl" nodal superconductor because the BdG nodal point is fourfold degenerate, which is equivalent to two distinct physical fermion degrees of freedom when the artificial Nambu doubling is taken away. This should not be confused with a Weyl (semi)metal for the following reasons: First, the nodal point is at $\bf k=0$, which is a time-reversal invariant momentum, and the particle-hole symmetry requires the Chern number of the BdG bands around the nodal point to vanish. Second, the BdG Hamiltonian cannot simply be a Nambu doubling of a Weyl (semi)metal because there is no regularizable charge preserving model with only one Weyl species.

\begin{figure}[htbp]
\centering\includegraphics[width=0.25\textwidth]{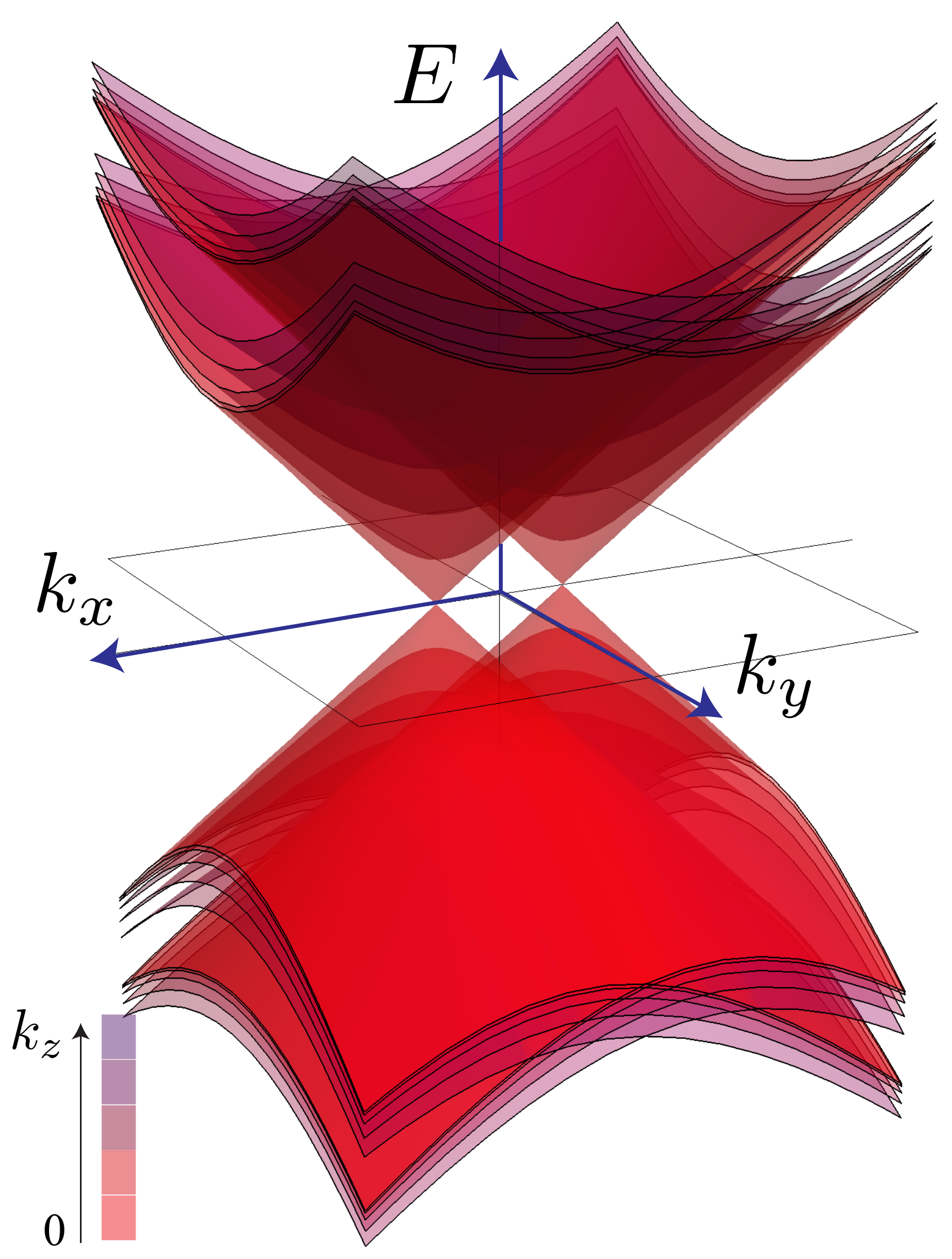}
\caption{The splitting of the two nodal Weyl points of a Dirac nodal superconductor by the symmetry preserving perturbation in Eq. \eqref{DiracSCsplit}.}\label{fig:Hcontsplit}
\end{figure}

A Weyl nodal superconductor is stable to all perturbations because there is no symmetry preserving gapping potentials and the nodal point is pinned by time-reversal. On the other hand, a Dirac nodal superconductor is only stable up to weak perturbation. While it is true that there is no symmetry preserving potentials that can immediately introduce an excitation energy gap, the two nodal Weyl species can be split in momentum space. For example, the following perturbation introduces the split. \begin{align}H_{\mathrm{SC-Dirac}}=\hbar v{\bf k}\cdot\vec{s}\mu_z\tau_z+us_x\mu_y\tau_x,\label{DiracSCsplit}\end{align} where the two nodal Weyl points split along the $x$-axis (See Figure~\ref{fig:Hcontsplit}). Each nodal Weyl point is fourfold degenerate. The net Chern number of the BdG bands around each nodal point is still trivial due to the glide symmetry. Instead, it is protected by the mirror winding number $N=1$, which is well defined as long as one stay in one of the two glide eigenspaces. However, if the perturbation is big enough, the two nodal Weyl points can be pushed to the boundary of the Brillouin zone, where the glide eigenvalues switch. The two nodal Weyl points will now have opposite mirror windings and can pair annihilate. A Dirac nodal superconductor is therefore stable against symmetry-preserving perturbations up to $u\lesssim hv/a$, where $a$ is the microscopic length scale of the glide translation.

In a similar way as the continuum model, the coupled wire model in Eq. \eqref{Eq:WeylHam} has two separated Weyl nodes at $K_1$ and $K_2$ (see Figure~\ref{fig:3Darray}(b)). They are pinned by the (anti-ferromagnetic) time-reversal symmetry and therefore the model is stable to single-body symmetry-preserving perturbations to arbitrary strength. If we dress the model with additional Dirac fermion flavors $d^1,\ldots,d^N$, there will be $N$ nodal Weyl points at each of the two high symmetry momenta $K_1$ and $K_2$. However, similar to the continuous case in Eq. \eqref{DiracSCsplit}, they can now split in pairs. For example, We here illustrate the case when there are $N=2$ fermion flavors. The unperturbed Hamiltonian $\mathcal{H}=\mathcal{H}^{(1)}+\mathcal{H}^{(2)}$ is two decoupled copies of the primitive one, where $\mathcal{H}^{(a)}$ is identical to Eq. \eqref{Eq:WeylHam} by substituting the fermions $R,L$ with $R^a,L^a$. We decompose the Dirac fermions $d=R,L$ into Majorana components $d^a=(\gamma^a+i\delta^a)/\sqrt{2}$ and introduce the symmetry-preserving dimerization \begin{align}\mathcal{H}_{\mathrm{dimer}}=iu\sum_{xy}\gamma^1_{x,y}\gamma^2_{x,y+1}+\delta^1_{x,y}\delta^2_{x,y+1}.\label{Hdimer}\end{align} Then, the full BdG Hamiltonian can be written as,
\begin{align}H_{BdG}^{N=2}({\bf k})=\begin{pmatrix}H_{BdG}({\bf k})&H_{\mathrm{dimer}}({\bf k})\\H_{\mathrm{dimer}}({\bf k})^\dagger&H_{BdG}({\bf k})\end{pmatrix},\label{dimerBdG}
\end{align}
where $H_{BdG}({\bf k})$ is the original $N=1$ Hamiltonian given in Eq. \eqref{Eq:BdG} and the off-diagonal term \begin{align}H_{\mathrm{dimer}}({\bf k})=\frac{iu}{2}\begin{pmatrix}0&1&0&0\\e^{i(k_1+k_2)}&0&0&0\\0&0&0&-e^{i(k_1+k_2)}\\0&0&-1&0\end{pmatrix}\end{align} dimerizes between the two flavors.

\begin{figure}[htbp]
\centering\includegraphics[width=0.45\textwidth]{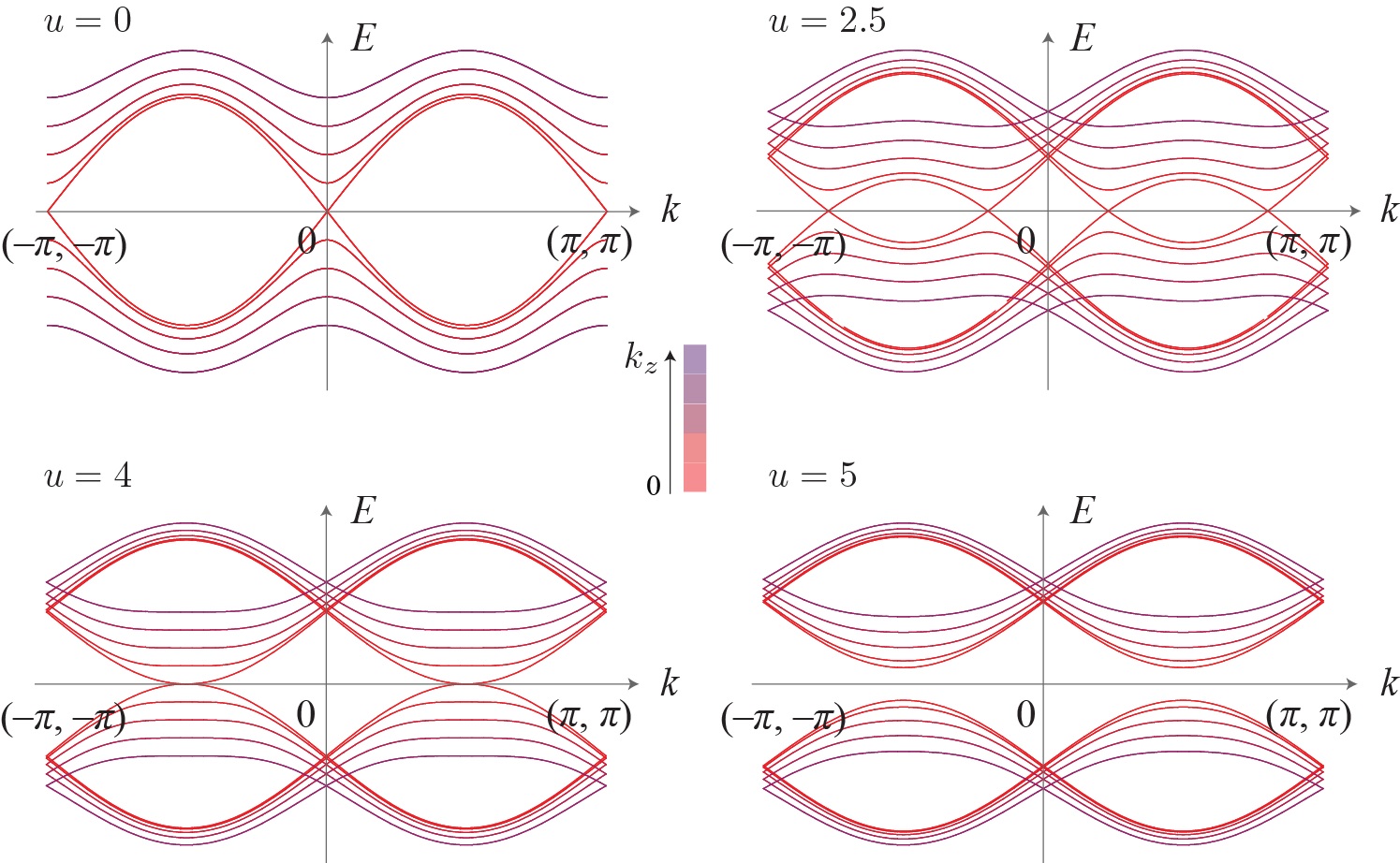}
\caption{The BdG energy spectrum of the dimerized system in Eq. \eqref{dimerBdG} along the $k_y$-direction (i.e.~$k_1=k_2$) for dimerization strength $u=0,2.5,4,5$ and $t_1=t_2=1$.}\label{fig:Z2bands}
\centering\includegraphics[width=0.48\textwidth]{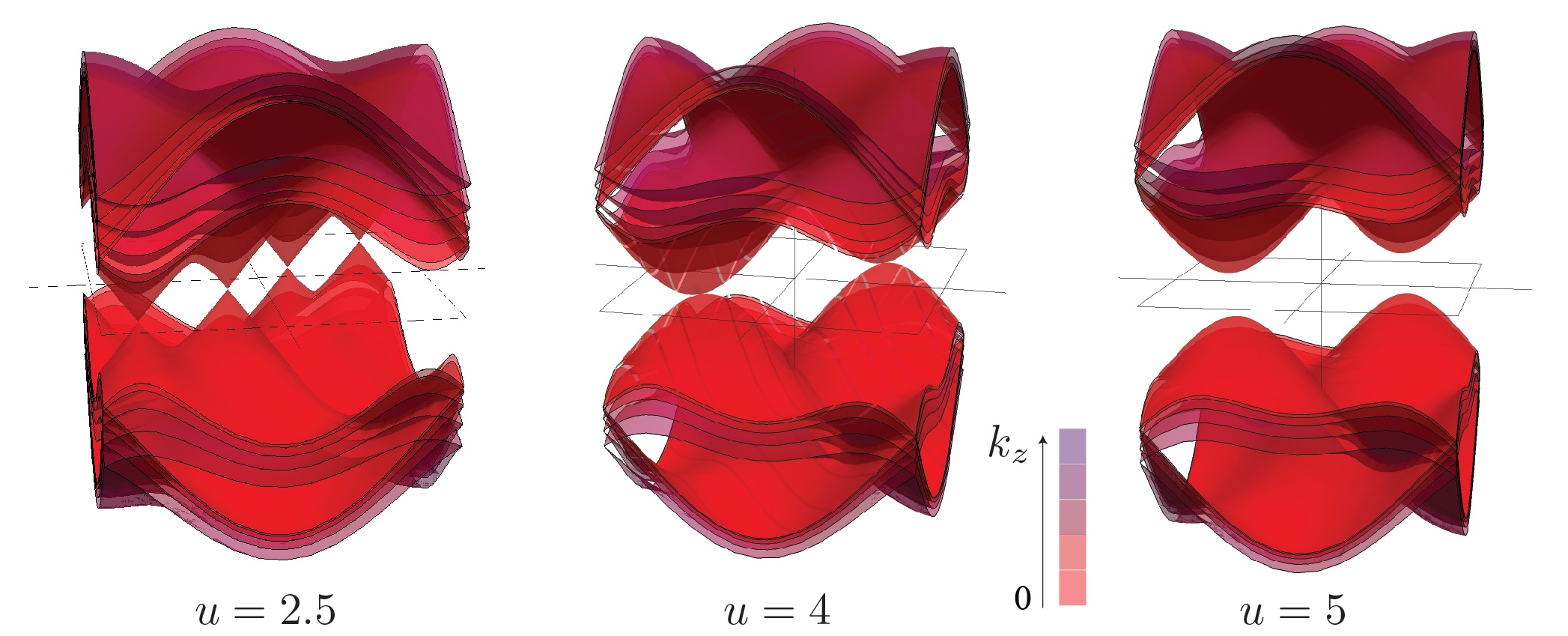}
\caption{The BdG energy spectrum of the dimerized system in Eq. \eqref{dimerBdG} over the entire Brillouin zone. As the two nodal Weyl points split in the momentum space, they meet with the other Weyl nodes with the opposite chirality and pair-annihilate each other. This insulator transition confirms the $\mathbb{Z}_2$ classification of our model.}\label{fig:Z2spectrum}
\end{figure}

Figure~\ref{fig:Z2bands} and \ref{fig:Z2spectrum} shows the dimerized energy spectrum. For small $u$, the nodal Weyl points pairwise split along the diagonal $k_y$-axis where $k_1=k_2$. The system remains gapless until the Weyl points originated from opposite momenta $K_{1,2}$ meet. This happens in a system with $u=4t_1=4t_2$ at $(k_1,k_2)=(\pi/2,\pi/2)$ and $(-\pi/2,-\pi/2)$, where Weyl points with opposite mirror windings pair up into quadratic band touching. A finite excitation energy gap opens when $u>4t_1=4t_2$. In the general situation, the coupled wire model with $N$ flavors is stable against any perturbation with strength $u\lesssim t_1,t_2$. If $N$ is odd, there is always an odd number of nodal Weyl points at each of the high symmetry momenta $K_{1,2}$ due to time reversal. They are robust against all single-body perturbations with arbitrary strength. As a result, the nodal coupled wire models are therefore $\mathbb{Z}$ classified for weak perturbation and $\mathbb{Z}_2$ classified for strong ones.
\section{Symmetry preserving many-body gapping interactions}\label{sec:manybody1}
In the previous section, we discussed the Dirac nodal superconductor under the single-body BCS mean-field description. In the low-energy limit, the nodal system was captured by the coupled-wire model in Eq. \eqref{Eq:WeylHam}, which exhibited a pair of massless Weyl fermions located at two time reversal invariant momenta $K_{1,2}$. The massless Weyl fermions are protected by the antiferromagnetic time-reversal $T_A$, mirror-glide $G$ as well as lattice translation $\mathsf{t}_{11},\mathsf{t}_{\bar{1}1}$ symmetries in Eq. \eqref{CWsymmetries}. In general, a nodal point is $\mathbb{Z}$-classified by the mirror-glide winding number in Eq. \eqref{Mwinding}, which counts the number (or net handedness) of massless Weyl fermions. This may be constructed by stacking $N$ copies of the fundamental model in Eq. \eqref{Eq:WeylHam}. This means that each vortex line now hosts, in general, a number of co-propagating chiral Dirac fermions $d^a=R^a$ or $L^a$, where $a$ is the flavor index that runs from 1 to $N$. The massless Weyl fermions are stable against the single-body symmetry preserving perturbations until they pair annihilate. If $N$ is odd, the anitferromagnetic time-reversal symmetry pins at least one massless Weyl node at each of the two high symmetry $K$ points in the Brillouin zone. As the Weyl node cannot move, they are robust even against pair annihilation. Under this non-interacting setup, in this section, we are interested in finding many-body interactions that introduces a finite excitation energy gap, for both even and odd flavor number $N$, while preserving the antiferromagnetic time-reversal, mirror-glide and lattice translation symmetries.

\begin{figure}[htbp]
\centering\includegraphics[width=0.45\textwidth]{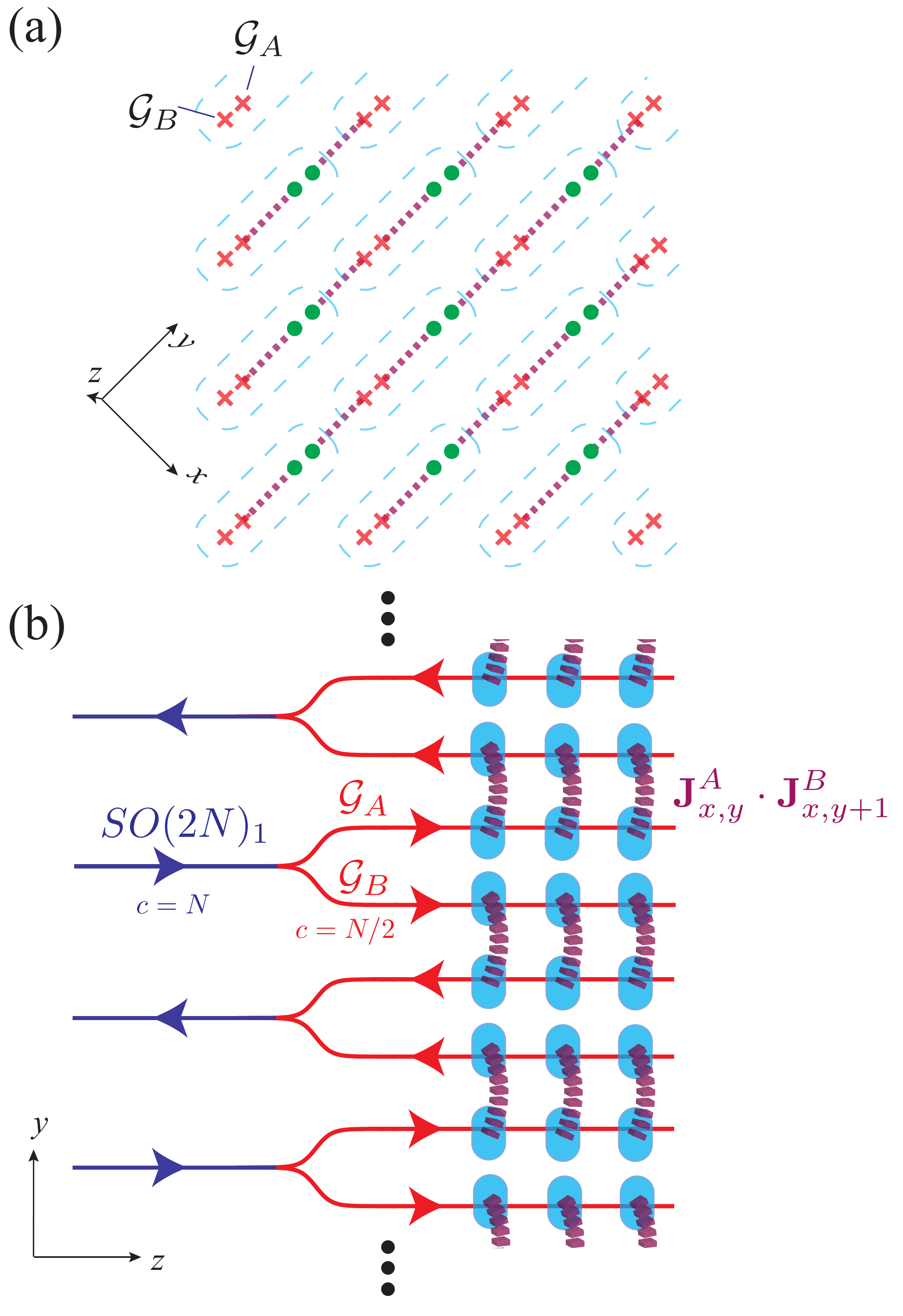}
\caption{Schematic representation of the many-body gapping interactions by interwire current backscatterings. The $N$ Dirac fermions per wire are divided into a pair of $\mathcal{G}$ affine Kac-Moody current algebras. (a) shows the current backscatterings in the 3D system viewed along the wire $z$-direction. (b) shows the current backscatterings on a single $yz$ layer.}\label{fig:gappingpotential}
\end{figure}

Our strategy is based on the coupled-wire construction presented in ref.~\onlinecite{PhysRevB.94.165142}. It relies on a bipartition of degrees of freedom on each vortex line. The degrees of freedom of the $2N$ chiral Majorana fermions $\gamma^a,\delta^a$, which pair into the Dirac fermions $d^a=(\gamma^a+i\delta^a)/\sqrt{2}$ for $a=1,\ldots,N$, are summarized by the $SO(2N)$ Kac-Moody current algebra at level 1 (also referred to as affine Lie algebra or Wess-Zumino-Witten theory). The strategy is to decompose the current algebra into a pair of identical and decoupled components \begin{align}SO(2N)_1\sim\mathcal{G}_A\times\mathcal{G}_B,\label{ABdecomposition}\end{align} where $\mathcal{G}_{A,B}$ are also affine Kac-Moody algebras and they act on decoupled Hilbert spaces $\mathcal{H}_{A,B}$. The decomposition in Eq. \eqref{ABdecomposition} is also known as conformal embedding in the conformal field theory context~\cite{bigyellowbook}. The current operators ${\bf J}^{A,B}$ in $\mathcal{G}_{A,B}$ can be expressed as combinations of products of the chiral fermions. For the most part in this paper, except for the $E_8$ algebra that we introduce in chapter \ref{Chapter:E8}, the current operators are fermion bilinears. For example, the $SO(\mathcal{N})_1$ current operators for a $(1+1)$D system of $\mathcal{N}$ chiral Majorana fermions $\psi^1,\ldots,\psi^{\mathcal{N}}$ are \begin{align}J_{ab}=i\psi^a\psi^b\label{SONcurrentdef}\end{align} for $1\leq a<b\leq\mathcal{N}$. Therefore, the many-body interactions, being two-body for the most part in this paper, are constructed by backscattering the $A$ and $B$ currents to neighboring vortex lines that counter-propagate in opposite directions, for example in the $y$ and $-y$ directions. \begin{align}\mathcal{H}_{\mathrm{int}}=u\sum_{xy}{\bf J}^A_{x,y}\cdot{\bf J}^B_{x,y+1},\label{Hint0}\end{align} where $u$ is the interaction strength. Figure~\ref{fig:gappingpotential} shows the schematic representation of the decomposition and the interwire backscattering terms. As the $A$ and $B$ sectors are decoupled, the backscattering terms do not compete and, in strong-coupling, they introduce a finite many-body energy gap. Moreover, the decomposition in Eq. \eqref{ABdecomposition} will be designed explicitly in a way so that the current backscattering interactions in Eq. \eqref{Hint0} preserve all of the symmetries present in the non-interacting construction.

The partition scheme in Eq. \eqref{ABdecomposition} is separated into two cases depending on whether the number $N$ of chiral Dirac fermions per line is even or odd. When $N=2r$ is even, the current algebra can be split in a pair of Dirac bundles \begin{align}SO(2N)_1\sim SO(2r)_1^A\times SO(2r)_1^B\label{evendecomposition}\end{align} that split the Dirac fermions into two groups $d^1,\ldots,d^r$ and $d^{r+1},\ldots,d^{2r}$. When $N=2r+1$, it may be tempting to decompose the chiral Majorana, $\gamma^a$ and $\delta^a$ into two groups. However, as the $\gamma$s and $\delta$s transform differently under the aforementioned symmetries, there is no bipartition such that $SO(2N)_1\sim SO(2r+1)_1^A\times SO(2r-1)_1^B$ that leads to symmetry-preserving backscattering interactions. We will focus on the special case when $N=9=3\times3$. Here, there is an alternative way of decomposing the current algebra \begin{align}SO(18)_1&\sim SO(9)_1^\gamma\times SO(9)_1^\delta\nonumber\\&\sim\left[SO(3)_3^{\gamma,A}\times SO(3)_3^{\delta,A}\right]\nonumber\\&\;\;\;\;\times\left[SO(3)_3^{\gamma,B}\times SO(3)_3^{\delta,B}\right],\label{N=9decomposition}\end{align} using the level-rank duality~\cite{bigyellowbook} that in general relates $SO(nk)_1\sim SO(n)_k\times SO(k)_n$. In the general odd case, one can always write $N=9+2r$ and decompose $SO(2N)_1\sim SO(18)_1\times SO(4r)_1$ first. The bipartition can then be performed individually for $SO(18)_1$ and $SO(4r)_1$.

The decompositions in Eq. \eqref{evendecomposition} and \eqref{N=9decomposition} lead to gapping interactions in Eq. \eqref{Hint0} that support fractional quasiparticle excitations. These are non-local excitations that are not integral combinations of local electronic BdG quasiparticles. Since the many-body interactions in Eq. \eqref{Hint0} have a layered structure and act within $y-z$ planes, these non-local quasiparticles are confined in two dimensions and can exhibit anyonic statistics. Additional interlayer condensation couplings may promote the layered systems into three dimensional topological phases that support quasi-string excitations and topological ground state degeneracies although the discussion of topological orders in three dimensions is out of the scope of this article. In addition, we will address an alternative set of many-body gapping interactions, when $N$ is a integer multiple of 16, that only allows local electronic BdG excitations. This stems from the decomposition \begin{align}SO(32)_1\sim(E_8)_1\times(E_8)_1\end{align} using the exceptional affine Lie algebra $E_8$ at level 1. The non-topologically ordered many-body gapping at $N=16$ reduces the $\mathbb{Z}$ classification of Dirac nodal superconductors to $\mathbb{Z}_{16}$, which resembles the cyclic classification of topological superconductors~\cite{LukaszChenVishwanath,MetlitskiFidkowskiChenVishwanath14}.

\subsection{The \texorpdfstring{$N=2r$}{N=2r} - Even Case}
We begin our discussion of the {$N=2r$} by expressing the Dirac fermion $d^a_{x,y}(z)=(\gamma^a_{x,y}(z)+i\delta^a_{x,y}(z))/\sqrt{2}\sim e^{i\phi^a_{x,y}(z)}$ as a vertex operator of the bosonized variable $\phi^a_{x,y}$. The kinetic Lagrangian density is \begin{gather}\mathcal{L}=\frac{1}{2\pi}\sum_{xy}(-1)^{x+y}\delta_{ab}\partial_t\phi^a_{x,y}\partial_z\phi^b_{x,y}-V_{ab}\partial_z\phi^a\partial_z\phi^b,\label{L0}\end{gather} where $V_{ab}$ is a non-universal velocity matrix. The bosonized variables obey the equal-time commutation relation \begin{align}
&
\left[\phi_{x,y}^a(z),\phi_{x',y'}^{a'}(z')\right]
\nonumber\\&
=i\pi(-1)^{\mathrm{min}\{x,x'\}+\mathrm{min}\{y,y'\}}\big[\delta_{xx'}\delta_{yy'}\delta_{aa'}\mathrm{sgn}(z'-z)
\nonumber\\&\;\;\;+\sigma_z\delta_{aa'}\delta_{yy'}\mathrm{sgn}(x-x')
\nonumber\\&\;\;\;+\sigma_z\delta_{yy'}\mathrm{sgn}(a-a')-\sigma_z\mathrm{sgn}(y-y')\big].
\label{ETCR}
\end{align} where $\mathrm{sgn}(s)=s/|s|$ if $s\neq0$ or $0$ if $s=0$. Here $\sigma_z$ is an auxiliary factor that anticommutes with the non-local mirror-glide operator, $G\sigma_zG^{-1}=-\sigma_z$. See Appendix \ref{Appendix:B} for the detailed explanation of the commutation relation. The bosonized variables transform under the lattice translations $\mathsf{t}_{11},\mathsf{t}_{\bar{1}1}$, antiferromagnetic time-reversal $T_A$ and mirror-glide $G$ symmetries according to \begin{gather}\mathsf{t}_{11}\phi^a_{x,y}(z)\mathsf{t}_{11}^{-1}=\phi^a_{x+1,y+1}(z),\nonumber\\\mathsf{t}_{\bar{1}1}\phi^a_{x,y}(z)\mathsf{t}_{\bar{1}1}^{-1}=\phi^a_{x-1,y+1}(z),\nonumber\\T_A\phi_{x,y}^a(z)T_A^{-1}=-\phi_{x,y+1}^a(z)+\frac{1-(-1)^{x+y}}{2}\pi,\nonumber\\G\phi^a_{x,y}(z)G^{-1}=-\phi_{x+1,y}^a(-z)+\frac{\pi}{2}.\label{bosonsymmtrans}\end{gather} These transformations are based on the symmetry operations performed on the chiral Dirac fermions in Eq. \eqref{CWsymmetries}. They are consistent with the equal-time commutation relation in Eq. \eqref{ETCR} and the algebraic relations $T_A^2=(-1)^F\mathsf{t}_{11}\mathsf{t}_{\bar{1}1}$, $G^2=\mathsf{t}_{11}\mathsf{t}_{\bar{1}1}^{-1}$ and $[\mathcal{O},\mathcal{O}']=0$ for $\mathcal{O},\mathcal{O}'=T_A,G,\mathsf{t}_{11},\mathsf{t}_{\bar{1}1}$, where $(-1)^F$ is the fermion parity operator and $T_A$ is antiunitary.

We split the $2r$ Dirac fermions per wire into two groups $d^1,\ldots,d^r$ and $d^{r+1},\ldots,d^{2r}$. Each group generates a $SO(2r)_1$ affine Kac-Moody algebra. We label the first by $SO(2r)_1^A$ and the second by $SO(2r)_1^B$. We now review and illustrate the (complexified) current operators. Using the bosonized variables, each of the two current algebras consists of $r$ Cartan generators $H^{A,a}=i\partial\phi^a$ and $H^{B,a}=i\partial\phi^{r+a}$, for $a=1,\ldots,r$. There are $2r(r-1)$ roots for each sector, \begin{align}E^{A,\boldsymbol\alpha}=e^{i\boldsymbol\alpha\cdot\boldsymbol\phi^A},\quad E^{B,\boldsymbol\alpha}=e^{i\boldsymbol\alpha\cdot\boldsymbol\phi^B}\end{align} where $\boldsymbol\phi^A=(\phi^1,\ldots,\phi^r)$ amd $\boldsymbol\phi^B=(\phi^{r+1},\ldots,\phi^{2r})$. $\boldsymbol\alpha$ is a $r$-dimensional vector with integral entries and length $|\boldsymbol\alpha|=\sqrt{2}$. In other words, each root vector has two and only two non-zero entries, each being $\pm1$. The interwire current backscattering in Eq. \eqref{Hint0} becomes \begin{align}\mathcal{H}_{\mathrm{int}}&=-u\sum_{xy}\sum_{\boldsymbol\alpha}E^{A,\boldsymbol\alpha}_{x,y}E^{B,\boldsymbol\alpha}_{x,y+1}\nonumber\\&=-u\sum_{xy}\sum_{\boldsymbol\alpha}\cos\left[\boldsymbol\alpha\cdot(\boldsymbol\phi^A_{x,y}+\boldsymbol\phi^B_{x,y+1})\right],\label{Hinteven}\end{align} where we have suppressed the terms involving the Cartan generators \begin{align}-u\sum_{xy}\sum_{a=1}^rH^{A,a}_{x,y}H^{B,a}_{x,y+1}=u\sum_{xy}\partial_x\boldsymbol\phi^A_{x,y}\cdot\partial_x\boldsymbol\phi^B_{x,y+1}\label{evenforwardscattering}\end{align} which only renormalizes the velocities in Eq. \eqref{L0}.

The sine-Gordon angle parameters \begin{align}\Theta^{\boldsymbol\alpha}_{x,y+1/2}=\boldsymbol\alpha\cdot(\boldsymbol\phi^A_{x,y}+\boldsymbol\phi^B_{x,y+1})\end{align} satisfy the ``Haldane's nullity condition~\cite{PhysRevLett.74.2090}" \begin{align}
\left[\Theta^{\boldsymbol\alpha}_{x,y+1/2}(z),\Theta^{\boldsymbol\alpha'}_{x',y'+1/2}(z')\right]=0.
\label{Haldanenulity}
\end{align} Since all root vectors $\boldsymbol\alpha$ are integral combinations of the $r$ simple roots, which is explicitly given as, \begin{align}\begin{pmatrix}|&\ldots&|\\\boldsymbol\alpha_1&\ldots&\boldsymbol\alpha_r\\|&\ldots&|\end{pmatrix}=\begin{pmatrix}1&0&\ldots&0&0\\-1&1&\ldots&0&0\\0&-1&\ldots&0&0\\\vdots&\vdots&\ddots&\vdots&\vdots\\0&0&\ldots&1&1\\0&0&\ldots&-1&1\end{pmatrix}_{r\times r},\end{align} there are only $r$ linearly independent angle variables $\Theta^{\boldsymbol\alpha}_{x,y+1/2}$ given a fixed $x,y$. In a periodic geometry $y=y+L$ for $L$ even, there are $rL$ independent sine-Gordon angle variables and the same number of counter-propagating pairs of neutral Dirac modes in a fixed layer, $x$. Moreover, assuming $u>0$, the sine-Gordon potential in Eq. \eqref{Hinteven} pins the uniform ($z$-independent) ground state expectation values $\langle\Theta^{\boldsymbol\alpha}_{x,y+1/2}(z)\rangle$ to be integer multiples of $2\pi$ for all $\boldsymbol\alpha$. This means the order parameters $\langle\Theta^{\boldsymbol\alpha}_{x,y+1/2}(z)\rangle$, although being linearly dependent, are not competing because an integral combination of integers is still an integer. We can therefore conclude that Eq. \eqref{Hinteven} introduces a finite excitation energy gap in the bulk.

It is straightforward to check that the gapping potential in Eq. \eqref{Hinteven} is symmetric under all the symmetries in defined in Eq. \eqref{bosonsymmtrans}. As the model in Eq. \eqref{Hinteven} is exactly solvable, the ground state must also preserve all symmetries. In fact, the symmetries in Eq. \eqref{bosonsymmtrans} require the angle order parameters to obey the following: \begin{align}\left\langle\Theta^{\boldsymbol\alpha}_{x,y+1/2}(z)\right\rangle&=-\left\langle\Theta^{\boldsymbol\alpha}_{x,y+3/2}(z)\right\rangle+\pi\boldsymbol\alpha\cdot{\bf t}\nonumber\\&=-\left\langle\Theta^{\boldsymbol\alpha}_{x+1,y+1/2}(-z)\right\rangle+\pi\boldsymbol\alpha\cdot{\bf t}\label{evenorderparameter}\\&=\left\langle\Theta^{\boldsymbol\alpha}_{x+1,y+3/2}(z)\right\rangle=\left\langle\Theta^{\boldsymbol\alpha}_{x-1,y+3/2}(z)\right\rangle,\nonumber\end{align} where ${\bf t}=(1,1,\ldots,1)^T$ and $\boldsymbol\alpha\cdot{\bf t}=\sum_{a=1}^r\alpha^a=2,0,-2$. We notice in passing that these are not the most primitive angle order parameters. For example, the vector and spinor fields of $SO(2r)_1$ correspond to smaller angle order parameters~\cite{PhysRevB.94.165142} \begin{align}\begin{array}{*{20}l}\Theta^a_{x,y+1/2}=\phi^a_{x,y}+\phi^{r+a}_{x,y+1}\\\Theta^{\boldsymbol\varepsilon}_{x,y+1/2}=\boldsymbol\varepsilon\cdot\left(\boldsymbol\phi^A_{x,y}+\boldsymbol\phi^B_{x,y+1}\right)/2\end{array}\end{align} respectively, where $\boldsymbol\varepsilon=(\varepsilon^1,\ldots,\varepsilon^r)^T$ for $\varepsilon^a=\pm1$. However, since these terms are not necessary in the discussion of the gapping potential, we will omit them.

Lastly, we express the gapping potential in Eq. \eqref{Hinteven}, including forward scattering terms expressed in Eq. \eqref{evenforwardscattering}, in terms of the Majorana fermions \begin{align}\mathcal{H}_{\mathrm{int}}&=2u\sum_{1\leq a_1<a_2\leq r}\left(\gamma_{x,y}^{a_1}\gamma_{x,y}^{a_2}\gamma_{x,y+1}^{r+a_1}\gamma_{x,y+1}^{r+a_2}\right.\nonumber\\&\quad\quad\quad\quad\quad\quad\left.+\delta_{x,y}^{a_1}\delta_{x,y}^{a_2}\delta_{x,y+1}^{r+a_1}\delta_{x,y+1}^{r+a_2}\right)\nonumber\\&\;\;\;-2u\sum_{1\leq a_1<a_2\leq r}\left(\gamma_{x,y}^{a_1}\delta_{x,y}^{a_2}\gamma_{x,y+1}^{r+a_1}\delta_{x,y+1}^{r+a_2}\right.\nonumber\\&\quad\quad\quad\quad\quad\quad\left.+\delta_{x,y}^{a_1}\gamma_{x,y}^{a_2}\delta_{x,y+1}^{r+a_1}\gamma_{x,y+1}^{r+a_2}\right),\label{Hinteven2}\end{align} where $d^a_{xy}=(\gamma^a_{xy}+i\delta^a_{xy})/\sqrt{2}\sim e^{i\phi^a_{xy}}$, for $a=1,\ldots,N=2r$. The Majorana fermions transform under the given symmetries \eqref{CWsymmetries} according to the following: \begin{gather}\begin{split}T_A\gamma_{x,y}^a(z)T_A^{-1}&= (-1)^{x+y}\gamma_{x,y+1}^a(z)\\T_A\delta_{x,y}^a(z)T_A^{-1}&=(-1)^{x+y+1}\delta_{x,y+1}^a(z)\end{split},\label{majoranaAFTR}\\\begin{split}G\gamma_{x,y}^a(z)G^{-1}&=\delta_{x+1,y}^a(-z)\\G\delta_{x,y}^a(z)G^{-1}&=\gamma_{x+1,y}^a(-z)\end{split},\label{majoranaG}\\\begin{split}\mathsf{t}_{11}\psi_{x,y}^a(z)\mathsf{t}_{11}^{-1}&=\psi_{x+1,y+1}^a(z)\\\mathsf{t}_{\bar{1}1}\psi_{x,y}^a(z)\mathsf{t}_{\bar{1}1}^{-1}&=\psi_{x-1,y+1}^a(z)\end{split},\label{majoranaT}\end{gather} where $\psi=\gamma$ or $\delta$. The two-body Hamiltonian in Eq. \eqref{Hinteven2}, therefore, preserves all symmetries present in our construction. In fact, the symmetries are preserved individually for each of the two lines in Eq. \eqref{Hinteven2}, and any of the two lines alone can already introduce a finite energy gap. We consider both so that the Hamiltonian takes the full Kac-Moody current backscattering form in Eq. \eqref{Hint0}. The relative minus sign between the two lines comes from Eq. \eqref{Hinteven}, where the current backscatterings are designed to be $E^{A,\boldsymbol\alpha}_{x,y}E^{A,\boldsymbol\alpha}_{x,y+1}$ rather than $E^{A,\boldsymbol\alpha}_{x,y}E^{A,-\boldsymbol\alpha}_{x,y+1}$. This is to ensure the ground state expectation values $i\langle\gamma^a_{x,y}\gamma^{r+a}_{x,y+1}\rangle$ and $i\langle\delta^a_{x,y}\delta^{r+a}_{x,y+1}\rangle$ to have the same sign so that the mirror-glide symmetry is not spontaneously broken. In retrospect, this is not surprising because the number of fermion flavors here is even, and the nodal model can acquire a single-body energy gap if the single-body potential is strong enough to pull the massless Weyl nodes together. The single-body potential that achieves this has already been given by the dimerization term $\mathcal{H}_{\mathrm{dimer}}$ in Eq. \eqref{Hdimer} when $N=2$, however, this splitting term can also be generalized for an arbitrary even $N$. It is not a coincidence that $\mathcal{H}_{\mathrm{dimer}}$ also pins the same ground state expectation values $i\langle\gamma^a_{x,y}\gamma^{r+a}_{x,y+1}\rangle$ and $i\langle\delta^a_{x,y}\delta^{r+a}_{x,y+1}\rangle$. This is because we can view the single-body Hamiltonian $\mathcal{H}_{\mathrm{dimer}}$ as the mean-field approximation of the two-body Hamiltonian in Eq. \eqref{Hinteven2}.

\subsection{The \texorpdfstring{$N=2r+1$}{N=2r+1} odd case}\label{sec:SO3}
From the previous discussion in section~\ref{sec:coupled}, we have seen that the nodal coupled wire model is stable to all single-body symmetry-preserving perturbations to arbitrary strength when the number of fermion flavors is odd. The focus of this section is to design exactly solvable two-body interactions that introduce a symmetry-preserving energy gap in the coupled wire construction when the number of fermion flavors is odd. We begin again by decomposing each Dirac fermion into a pair of Majorana modes $d^a_{xy}(z)=(\gamma^a_{xy}(z)+i\delta^a_{xy}(z))/\sqrt{2}$, where $a=1,\ldots,N=2r+1$ is the fermion flavor label. Additionally, we assume that the Majorana operators transform identically as defined in Eq. \eqref{majoranaAFTR}, \eqref{majoranaG} and \eqref{majoranaT}.

At this point, it may be tempting to group the $2N$ Majoranas per wire into two collections, namely $\psi^1,\ldots,\psi^N$ and $\psi^{N+1},\ldots,\psi^{2N}$, and consider the $SO(N)$ current backscatterings such as \begin{align}\mathcal{H}=\sum_{xy}\sum_{1\leq a<b\leq N}u_{ab}\psi^a_{x,y}\psi^b_{x,y}\psi^{N+a}_{x,y+1}\psi^{N+b}_{x,y+1}.\label{Hfail}\end{align} However, because the numbers of $\gamma$'s and $\delta$'s are odd, there must be an imbalance in the number of $\gamma$ and $\delta$ in each of the two collections. Consequently, there is no biparition of Majorana fermions that is compatible with the symmetries. This is because the antiferromagnetic time-reversal action in Eq. \eqref{majoranaAFTR} on both $\gamma$ and $\delta$ are different by a sign, while the mirror-glide action in Eq. \eqref{majoranaG} switches between $\gamma$ and $\delta$. In other words, there is no orthogonal basis transformation of fermions $(\gamma^1,\ldots,\gamma^N,\delta^1,\ldots,\delta^N)\leftrightarrow(\psi^1,\ldots,\psi^{2N})$ that achieves a symmetry-invariant bipartition $SO(N)_1^A=\langle\psi^1,\ldots,\psi^N\rangle$ and $SO(N)_1^B=\langle\psi^{N+1},\ldots,\psi^{2N}\rangle$ so that the symmetries are closed within each of the two sectors. Moreover, as seen in the previous section, there is no single-body symmetric gapping, nor any many-body Hamiltonian, such as Eq. \eqref{Hfail}, that admits a single-body mean-field solution must fail.

The construction of the two-body interactions that can accomplish a symmetric energy gap relies on another type of bipartition. First, we separate the Majorana fermions into \begin{align}SO(N)_1^{\gamma,\delta}\sim SO(9)_1^{\gamma,\delta}\times SO(2n)_1^{\gamma,\delta}\end{align} for both the $\gamma$ and $\delta$ ones, where $N=9+2n$. This can clearly be done when $N$ is not less than 9. The $SO(9)_1^\gamma$ sector is generated by $\gamma^1,\ldots,\gamma^9$ and the $SO(2n)_1^\gamma$ sector is generated by $\gamma^{10},\ldots,\gamma^N$. A similar decomposition applies to the $\delta$'s as well. If $N$ is smaller than 9, we extend the number of Dirac channels per wire by counter-propagating ones. This can be done by wire reconstruction that pulls $2n'=9-N$ counter-propagating pairs of Dirac modes to zero energy while keeping the net chirality $N=(N+2n')-2n'$, which is the difference of numbers of forward and backward moving Dirac fermions. The separation can now be done the same as before except the Majorana's in $SO(9)_1$ are forward propagating and the ones in $\overline{SO(2n')_1}$ are backward propagating. We also denote the counter-propagating $\overline{SO(2n')_1}$ sector by $SO(-2n')_1$.

The $SO(2n)_1^\gamma$ and $SO(2n)_1^\delta$ sectors can be gapped by either using the single-body dimerization in Eq. \eqref{Hdimer} or the two-body interaction in Eq. \eqref{Hinteven} described in the previous subsection. We now focus on the $SO(9)_1$ sectors, and without loss of generality, we now take $N=9$. Similar gapping potentials were presented in Ref.\onlinecite{PhysRevB.94.165142} in the context of topological superconducting surface states. This gapping potential relies on the splitting (also known as conformal embedding or level rank duality in the CFT context) \begin{align}SO(9)_1\supseteq SO(3)_3^A\times SO(3)_3^B\label{SO9SO3AB}\end{align} for both the $\gamma$ and $\delta$. We now apply this splitting to our case here by noting that for both the 9 $\gamma$'s and 9 $\delta$'s, we define two $SO(3)_3$ Kac-Moody current algebras \begin{align}\begin{split}J^{\psi,A}_{\mathsf{x}}=i(\psi_2\psi_3+\psi_5\psi_6+\psi_8\psi_9)\\J^{\psi,A}_{\mathsf{y}}=i(\psi_3\psi_1+\psi_6\psi_4+\psi_9\psi_7)\\J^{\psi,A}_{\mathsf{z}}=i(\psi_1\psi_2+\psi_4\psi_5+\psi_7\psi_8)\\J^{\psi,B}_{\mathsf{x}}=i(\psi_4\psi_7+\psi_5\psi_8+\psi_6\psi_9)\\J^{\psi,B}_{\mathsf{y}}=i(\psi_7\psi_1+\psi_8\psi_2+\psi_9\psi_3)\\J^{\psi,B}_{\mathsf{z}}=i(\psi_1\psi_4+\psi_2\psi_5+\psi_3\psi_6)\end{split},\label{SO33currentdef}\end{align} where $\psi=\gamma$ or $\delta$.

We now briefly summarize the conformal structures of the Kac-Moody current algebras. The details associated with these can be found in Ref.~\onlinecite{PhysRevB.94.165142} and will not be repeated here. The current operators obey the product expansion \begin{align}&J^{\psi,C}_{\mathsf{j}}(w)J^{\psi',C'}_{\mathsf{j}'}(w')\\&=\delta^{\psi\psi'}\delta^{CC'}\left[\frac{3\delta_{\mathsf{jj}'}}{(w-w')^2}+\frac{i\epsilon_{\mathsf{jj}'\mathsf{j}''}}{w-w'}J^\psi_{\mathsf{j}''}(w')\right]+\ldots\nonumber\end{align} where $w,w'\sim \tau+(-1)^{x+y}iz$ is the holomorphic/anti-holomorphic parameter, $C,C'=A,B$ and $\mathsf{j},\mathsf{j}',\mathsf{j}''=\mathsf{x},\mathsf{x},\mathsf{z}$. Here, $\epsilon_{\mathsf{jj}'\mathsf{j}''}$ is the structure factor of $SO(3)$, which is also the antisymmetric Levi-Civita tensor. The factor of 3 in the most singular piece sets the level of the Kac-Moody algebras. The four sectors $(\gamma,A)$, $(\gamma,B)$, $(\delta,A)$ and $(\delta,B)$ are completely decoupled from one another as mutual products are non-singular. This means that they act independently on orthogonal many-body Hilbert spaces. This is a non-trivial result because for each type of fermions $\psi=\gamma$ or $\delta$, both the $A$ and $B$ currents exhaust all 9 Majorana fermions. The separation of Hilbert spaces is, therefore, a non-trivial {\em fractionalization} beyond any fermionic mean-field approximation.

The embedding in Eq. \eqref{SO9SO3AB} is maximal in the sense that there are no degrees of freedom that remain unaccounted. This can be verified by the identification of the energy-momentum tensors \begin{align}T_{SO(3)_3^{\psi,A}}+T_{SO(3)_3^{\psi,B}}=T_{SO(9)_1^\psi}\label{EMtensoraddition}\end{align} for $\psi=\gamma,\delta$, where each tensor takes the Suguwara (normal ordered) representation~\cite{bigyellowbook} \begin{align}T_{SO(9)_1^\psi}&=\frac{1}{16}\sum_{a<b}J_{ab}^\psi J_{ab}^\psi=-\frac{1}{2}\sum_{a=1}^9\psi^a\partial\psi^a,\\T_{SO(3)_3^{\psi,C}}&=\frac{1}{8}\sum_{\mathsf{j}=\mathsf{x},\mathsf{y},\mathsf{z}}J^{\psi,C}_{\mathsf{j}}J^{\psi,C}_{\mathsf{j}}=-\frac{1}{4}\sum_{a=1}^9\psi^a\partial\psi^a-\hat{C},\\\hat{C}&=\pm\frac{1}{4}\left(\psi_{2356}+\psi_{2389}+\psi_{5689}+\psi_{1245}\right.\nonumber\\&\quad\quad\left.+\psi_{1278}+\psi_{4578}+\psi_{7182}+\psi_{7193}+\psi_{8293}\right),\nonumber\end{align} where the current operators $J_{ab}$ for $SO(9)_1$ was defined in Eq. \eqref{SONcurrentdef}, the sign of $\hat{C}$ is positive when $C=A$ or negative when $C=B$, and $\psi_{abcd}$ is the 4-fermion product $\psi_a\psi_b\psi_c\psi_d$. Each of the energy-momentum tensors satisfies the self-operator product expansion \begin{align}T(w)T(w')=\frac{c/2}{(w-w')^4}+\frac{2T(w')}{(w-w)^2}+\frac{\partial T(w')}{w-w'}+\ldots,\end{align} where the chiral central charge $c$ is $9/2$ for $SO(9)_1$ or $9/4$ for $SO(3)_3$. In particular, the identification in Eq. \eqref{EMtensoraddition} makes sure the chiral central charge is divided in equal parts through the conformal embedding in Eq. \eqref{SO9SO3AB}, i.e.~$9/2=9/4+9/4$. Moreover, mutual products between distinct $SO(3)_3^{\psi,C}$ sectors are non-singular due to the fact that the current algebras decouple.

\begin{figure}[htbp]
\centering\includegraphics[width=0.4\textwidth]{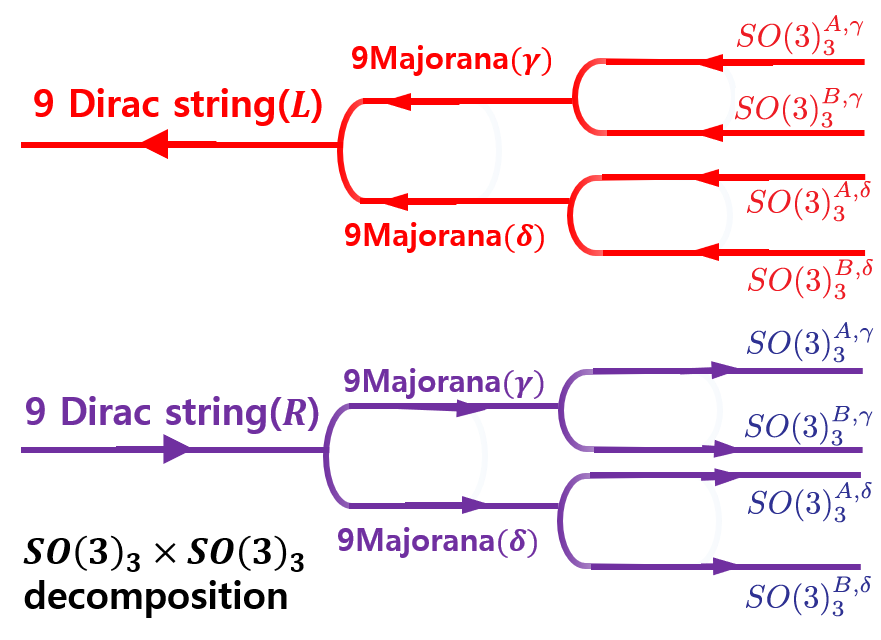}
\centering\includegraphics[width=0.4\textwidth]{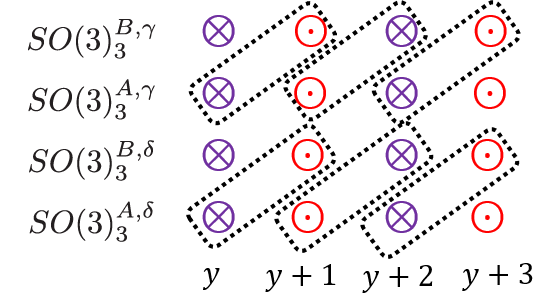}
\caption{(a) Schematic representations of the $SO(9)_1\supseteq SO(3)_3\times SO(3)_3$ decomposition. When $N=9$, we decompose each Dirac string into two Majorana fermions, $\gamma$ and $\delta$. Then each set of 9 Majorana fermions is decomposed into $SO(3)_3\times SO(3)_3$. (b) Schematic figure of the $SO(3)_3\times SO(3)_3$ many-body gapping potential. The boxes represent the current backscattering term between $SO(3)^{A,\psi}$ and $SO^{B,\psi}$ of adjacent wire in $y$ direction, for $\psi=\gamma,\delta$.}
\label{Fig:SO3}
\end{figure}

We now define the two-body potential using $SO(3)_3$ current backscatterings \begin{align}\mathcal{H}_{\mathrm{int}}=u\sum_{xy}{\bf J}^{\gamma,A}_{x,y}\cdot{\bf J}^{\gamma,B}_{x,y+1}+{\bf J}^{\delta,A}_{x,y}\cdot{\bf J}^{\delta,B}_{x,y+1},\label{HintSO3}\end{align} where ${\bf J}=(J_{\mathsf{x}},J_{\mathsf{y}},J_{\mathsf{z}})$ are the $SO(3)_3$ current operators defined in Eq. \eqref{SO33currentdef}. Figure \ref{Fig:SO3} shows the schematic figure of the gapping potential. From Eq. \eqref{majoranaAFTR}, \eqref{majoranaG} and \eqref{majoranaT}, we see that the potential preserves the antiferromagnetic time-reversal, glide and lattice translation symmetry. The gapping potential does not admit a mean-field solution with fermion bilinear order parameters $i\langle\psi^a\psi^b\rangle$ like those in Eq. \eqref{Hinteven2}. Therefore, to present the order parameters of the interaction in Eq. \eqref{HintSO3}, we introduce a further fractionalization (also known as coset construction~\cite{bigyellowbook} in the CFT context) for each of the four sectors $C=A,B$ and $\psi=\gamma,\delta$ \begin{align}SO(3)_3\sim SO(2)_3\times\mathbb{Z}_6,\label{Z6parafermion}\end{align} where $\mathbb{Z}_6$ represents the parafermion coset CFT~\cite{FateevZamolodchikov82,ZamolodchikovFateev85} $SO(3)_3/SO(2)_3=SU(2)_6/U(1)_6$.

The decomposition is done by first grouping three pairs of Majorana fermions into three neutral Dirac fermions in each sector \begin{align}f^{A,\psi}_1&=\frac{\psi^1+i\psi^2}{\sqrt{2}},&f^{B,\psi}_1&=\frac{\psi^1+i\psi^4}{\sqrt{2}}\nonumber\\f^{A,\psi}_2&=\frac{\psi^4+i\psi^5}{\sqrt{2}},&f^{B,\psi}_2&=\frac{\psi^2+i\psi^5}{\sqrt{2}}\nonumber\\f^{A,\psi}_3&=\frac{\psi^7+i\psi^8}{\sqrt{2}},&f^{B,\psi}_3&=\frac{\psi^3+i\psi^6}{\sqrt{2}}\nonumber\end{align} and bosonize \begin{align}f^{C,\psi}_j\sim\exp\left(i\tilde\phi^{C,\psi}_j\right)\label{so(3)3bosonization}\end{align} for $j=1,2,3$, $C=A,B$ and $\psi=\gamma,\delta$. The bosonic $SO(2)_3=U(1)_6$ sector is generated by the diagonal combination \begin{align}\Phi^{C,\psi}=\frac{\tilde\phi^{C,\psi}_1+\tilde\phi^{C,\psi}_2+\tilde\phi^{C,\psi}_3}{3}.\end{align} This leaves behind the orthogonal compliment $\phi^{C,\psi}_{\sigma,j}=\tilde\phi^{C,\psi}_j-\Phi^{C,\psi}$ and the Majorana fermions $\psi^3,\psi^6,\psi^9$ for the $A$ sector or $\psi^7,\psi^8,\psi^9$ for the $B$ sector. They combine into the parafermions \begin{align}\begin{split}\Psi^{A,\psi}&=\frac{1}{\sqrt{3}}\left(e^{i\phi_{\sigma,1}^{A,\psi}}\psi^3+e^{i\phi_{\sigma,2}^{A,\psi}}\psi^6+e^{i\phi_{\sigma,3}^{A,\psi}}\psi^9\right)\\\Psi^{B,\psi}&=\frac{1}{\sqrt{3}}\left(e^{i\phi_{\sigma,1}^{B,\psi}}\psi^7+e^{i\phi_{\sigma,2}^{B,\psi}}\psi^8+e^{i\phi_{\sigma,3}^{B,\psi}}\psi^9\right)\end{split},\label{Z6parafermiondefinition}\end{align} which generate the $\mathbb{Z}_6$ sector. The conformal field theory structures of the $SO(2)_3$ and $\mathbb{Z}_6$ are discussed in Ref.~\onlinecite{PhysRevB.94.165142} and will not be repeated here.

The coset construction in Eq. \eqref{Z6parafermion} allows the decomposition of the $SO(3)_3$ Kac-Moody current operators and consequently the potential in Eq. \eqref{HintSO3} \begin{align}\mathcal{H}_{\mathrm{int}}=3u\sum_{xy}\sum_{\psi=\gamma,\delta}e^{i(\Phi^{A,\psi}_{x,y}+\Phi^{B,\psi}_{x,y+1})}\Psi^{A,\psi}_{x,y+1}\Psi^{B,\psi}_{x,y+2}+h.c.,\label{HintZ6}\end{align}
where we have dropped the forward scatterings \begin{align}9u\sum_{xy}\sum_{\stackrel{C=A,B}{\psi=\gamma,\delta}}\partial_z\Phi^{C,\psi}_{x,y}\partial_z\Phi^{C,\psi}_{x,y+1}\end{align} from Eq. \eqref{HintSO3} that only renormalize the velocity for the boson in $SO(2)_3$. Order parameters are given by the ground state expectation values $\left\langle\Phi^{A,\psi}_{x,y}+\Phi^{B,\psi}_{x,y+1}\right\rangle$ and $\left\langle\Psi^{A,\psi}_{x,y}\Psi^{B,\psi}_{x,y+1}\right\rangle$, for $\psi=\gamma,\delta$. The symmetries defined in Eq. \eqref{majoranaAFTR}, \eqref{majoranaG} and \eqref{majoranaT} require \begin{align}\left\langle\Psi^{A,\psi}_{x,y}\Psi^{B,\psi}_{x,y+1}\right\rangle&=-\left\langle\Psi^{A,\psi}_{x,y+1}\Psi^{B,\psi}_{x,y+2}\right\rangle\nonumber\\&=-\left\langle\Psi^{A,\psi}_{x+1,y}\Psi^{B,\psi}_{x+1,y+1}\right\rangle\nonumber\\\left\langle\Psi^{A,\gamma}_{x,y}\Psi^{B,\gamma}_{x,y+1}\right\rangle&=-\left\langle\Psi^{A,\delta}_{x,y}\Psi^{B,\delta}_{x,y+1}\right\rangle\nonumber\\\left\langle\Psi^{A,\psi}_{x,y}\Psi^{B,\psi}_{x,y+1}\right\rangle&=-\left\langle\Psi^{A,\psi}_{x,y+1}\Psi^{B,\psi}_{x,y+2}\right\rangle\nonumber\\&=-\left\langle\Psi^{A,\psi}_{x+1,y}\Psi^{B,\psi}_{x+1,y+1}\right\rangle\nonumber\\\left\langle\Psi^{A,\gamma}_{x,y}\Psi^{B,\gamma}_{x,y+1}\right\rangle&=-\left\langle\Psi^{A,\delta}_{x,y}\Psi^{B,\delta}_{x,y+1}\right\rangle.\label{oddorderparameter}\end{align} Similar to Eq. \eqref{evenorderparameter} in the even case, the order parameters defined in Eq. \eqref{oddorderparameter} are not unique choices. For example, there are order parameters that correspond to non-Abelian twist fields in the $\mathbb{Z}_6$ sector that have quantum dimension greater than 1. However, they are not essential in the discussion of the symmetry-preserving gapping potential and are omitted.
\subsection{$E_8$ unimodular gapping potential}\label{Chapter:E8}
In the previous section, we found that even and odd copies of the $3$D Dirac nodal superconductor can be gapped out by many-body gapping potentials, which support non-trivial topological order. In this section, we now focus on the special case when there are $N=16$ copies of the Dirac nodal superconductor exists. In this case, we can construct a many-body gapping potential by utilizing $SO(32)\sim E_8\times E_8$ decomposition, where $E_8$ is the largest exceptional simple Lie algebra. The $E_8\times E_8$ decomposition is atypical from the previous decompositions since the roots of the $E_8$ Lie algebra form an even unimodular lattice. Here, we show that this property of the $E_8$ algebra allows us to construct many-body gapping potential in the $3D$ Dirac nodal superconductor that does not possess topological order.

Before going into the details on the construction of the gapping term, we briefly explain the $E_8$ Kac-Moody algebra at level one using bosonized variables. In addition to 8 Cartan generators $\partial\phi_I$, $I=1,\ldots,8$, the $(E_8)_1$ algebra is generated by the vertex operators $E^{\boldsymbol\alpha}=e^{i\boldsymbol\alpha\cdot\boldsymbol\phi}$, where $\boldsymbol\alpha$ is a root vector of the $E_8$ lattice. The $E_8$ lattice $\mathcal{L}_{E_8}$ is an 8 dimensional lattice generated by 8 simple root vectors. It is an even unimodular lattice in the sense that the norm square $|{\bf v}|^2$ of a lattice vector is even, and the dual lattice $\mathcal{L}_{E_8}^\ast$, which consists of dual vectors ${\bf v}^\ast$ whose scalar product with any $E_8$ lattice vector ${\bf v}$ is integral, is the $E_8$ lattice itself. In particular, there are 240 root vectors $\boldsymbol\alpha$ with norm square $|\boldsymbol\alpha|^2=2$ so that the vertex operators $E^{\boldsymbol\alpha}$ have unit spin and represent the $E_8$ Kac-Moody current.

The total 240 roots separate into two distinct sets. The first set consists of $112=C^8_2\times4$ roots of $SO(16)$ and the second set consists of $128=2^7$ even spinors. The conventional choice of roots embeds them in the 8 dimensional Euclidean space. The $SO(16)$ roots are taken to be integral vectors with two and only two non-zero components, each being $\pm1$. The corresponding vertex operators $E^{\boldsymbol\alpha}$ are fermion bilinears $d_ad_b$, $d_ad_b^\dagger$, $d_a^\dagger d_b$ and $d_a^\dagger d_b^\dagger$, for $1\leq a<b\leq8$. The even spinors are represented by half-integral vectors $\boldsymbol\epsilon/2=(\epsilon_1/2,\ldots,\epsilon_8/2)$, where $\epsilon_a=\pm1$, with overall positive sign $\epsilon_1\ldots\epsilon_8=1$. They corresponds to spinor vertex operators $e^{i\boldsymbol\epsilon\cdot\boldsymbol\phi/2}$, which are products of half fermions. Within the 240 roots, one can pick a set of 8 linearly independent simple roots that generate the entire set. \begin{gather}\begin{pmatrix}|&\ldots&|\\\boldsymbol\alpha_1&\ldots&\boldsymbol\alpha_8\\|&\ldots&|\end{pmatrix}=\left(\begin{smallmatrix}
 1 & 0 & 0 & 0 & 0 & 0 & 0 & -\frac{1}{2} \\
 -1 & 1 & 0 & 0 & 0 & 0 & 0 & -\frac{1}{2} \\
 0 & -1 & 1 & 0 & 0 & 0 & 0 & -\frac{1}{2} \\
 0 & 0 & -1 & 1 & 0 & 0 & 0 & -\frac{1}{2} \\
 0 & 0 & 0 & -1 & 1 & 0 & 0 & -\frac{1}{2} \\
 0 & 0 & 0 & 0 & 1 & -1 & 0 & -\frac{1}{2} \\
 0 & 0 & 0 & 0 & 0 & 1 & -1 & -\frac{1}{2} \\
 0 & 0 & 0 & 0 & 0 & 0 & 1 & -\frac{1}{2} \end{smallmatrix}\right).\label{E8simpleroots0}\end{gather} Their scalar products $\boldsymbol\alpha_I\cdot\boldsymbol\alpha_J=(K_{E_8})_{IJ}$ recover by the Cartan matrix of $E_8$ \begin{align}K_{E_8}=\left(\begin{smallmatrix}2&-1&&&&&&\\-1&2&-1&&&&&\\&-1&2&-1&&&&\\&&-1&2&-1&&&\\&&&-1&2&-1&&-1\\&&&&-1&2&-1&\\&&&&&-1&2&\\&&&&-1&&&2\end{smallmatrix}\right).\label{KE8}\end{align}

Unfortunately, the above conventional choice of roots involves spinors that are combinations of half fermions, which are non-local. In order to realize the $E_8$ algebra as integral combination of local fermions, we first extend the 8 chiral Dirac fermions by an additional counter-propagating pair of non-chiral Dirac fermions. This can be achieved by using the vortex reconstruction whereby we pull the addition non-chiral pair from high-energy to low-energy. The reconstruction does not alter the chirality $c=8=9-1$ of a vortex, which now consists of 9 forward propagating Dirac fermions and 1 backward propagating one. The $E_8$ lattice is now embedded in a $10=1+9$ dimensional ``Minkowski" space with metric $\eta=\mathrm{diag}(-1,1,\ldots,1)$. The $E_8$ roots consists of a subset of integral vectors ${\bf v}$ with norm square ${\bf v}^T\eta{\bf v}=-v_0^2+v_1^2+\ldots+v_9^2=2$. We begin with the roots of $SU(8)$, $\boldsymbol\alpha_{SU(8)}={\bf e}_a-{\bf e}_b$, where $1\leq a,b\leq8$ and $a\neq b$. These roots correspond to the fermion bilinear vertex operators $E^{\boldsymbol\alpha_{SU(8)}}\sim d_ad_b^\dagger$ and they obey the operator product expansion that defines the $SU(8)$ Kac-Moody algebra at level 1, \begin{align}E^{\boldsymbol\alpha_{SU(8)}}(w)E^{-\boldsymbol\alpha_{SU(8)}}(w')&=1/(w-w')^2+\ldots,\nonumber\\E^{\boldsymbol\alpha_{SU(8)}}(w)E^{\boldsymbol\alpha'_{SU(8)}}(w')&=c_{\alpha\alpha'}E^{\boldsymbol\alpha''_{SU(8)}}(w')/(w-w')\nonumber\\&\;\;\;+\ldots,\end{align} if $\boldsymbol\alpha''_{SU(8)}=\boldsymbol\alpha'_{SU(8)}+\boldsymbol\alpha'_{SU(8)}$, or non-singular if otherwise, where $w\sim\tau+iz$ is the complex space-time parameter and the cocycle factor $c_{\alpha\alpha'}$ is a scalar phase.

These 56 roots can be extended to $SO(16)$ by including two 28-dimensional irreducible representations of $SU(8)$. The first corresponds to 28 positive root vectors $\boldsymbol\alpha_{(+{\bf 28})}=2{\bf e}_0+n_1{\bf e}_1+\ldots+n_8{\bf e}_8$, where two of $n_1,\ldots,n_8$ are 0's and the rest are 1's. The second corresponds to 28 negative roots $\boldsymbol\alpha_{(-{\bf 28})}=-2{\bf e}_0-n_1{\bf e}_1-\ldots-n_8{\bf e}_8$. Each forms a super-selection sector that is closed under the $SU(8)$ Kac-Moody algebra \begin{align}E^{\boldsymbol\alpha_{SU(8)}}(w)E^{\boldsymbol\alpha'_{(\pm{\bf 28})}}(w')=c_{\alpha\alpha'}E^{\boldsymbol\alpha''_{(\pm{\bf 28})}}(w')/(w-w')+\ldots\end{align} if $\boldsymbol\alpha''_{(\pm{\bf 28})}=\boldsymbol\alpha_{SU(8)}+\boldsymbol\alpha'_{(\pm{\bf 28})}$, or non-singular if otherwise. The root vectors are chosen so that $\boldsymbol\alpha_{(\pm{\bf28})}^T\eta\boldsymbol\alpha_{(\pm{\bf28})}=2$ so that the vertices $E^{\boldsymbol\alpha_{(\pm{\bf28})}}$ have spin 1. We label $\boldsymbol\alpha_{SO(16)}$ to be the 112 roots for $SO(16)$, and they consists of $\boldsymbol\alpha_{SO(16)}$ and $\boldsymbol\alpha_{(\pm{\bf28})}$.

Next, the 112 $SO(16)$ roots can be extended to the full $E_8$ by including two 56-dimensional irreducible representations of $SU(8)$ and the two 8-dimensional vector representations of $SU(8)$. The first irreducible representation that we include is associated with the 56 positive root vectors $\boldsymbol\alpha_{(+{\bf56})}={\bf e}_0+m_1{\bf e}_1+\ldots+m_8{\bf e}_8$, where three of $n_1,\ldots,n_8$ are 1's and the rest are 0's. The conjugate representation is associated with the 56 negative roots $\boldsymbol\alpha_{(-{\bf56})}=-{\bf e}_0-m_1{\bf e}_1-\ldots-m_8{\bf e}_8$. The 8 positive $SU(8)$ vectors are $\boldsymbol\alpha_{(+{\bf8})}=3{\bf e}_0+v_1{\bf e}_1+\ldots+v_8{\bf e}_8$, where all but one $v_a=1$ and the remaining is 2. The 8 negative $SU(8)$ vectors are the conjugate $\boldsymbol\alpha_{(-{\bf8})}=-3{\bf e}_0-v_1{\bf e}_1-\ldots-v_8{\bf e}_8$. We label $\boldsymbol\alpha_{\mathrm{spinor}}$ to be the $128=56+56+8+8$ additional vectors $\boldsymbol\alpha_{(\pm{\bf56})}$ and $\boldsymbol\alpha_{(\pm{\bf8})}$ that constitute the even spinor representation for the $SO(16)$ algebra. \begin{align}E^{\boldsymbol\alpha_{SO(16)}}(w)E^{\boldsymbol\alpha'_{\mathrm{spinor}}}(w')&=c_{\alpha\alpha'}E^{\boldsymbol\alpha''_{\mathrm{spinor}}}(w')/(w-w')\nonumber\\&\;\;\;+\ldots\end{align} if $\boldsymbol\alpha''_{\mathrm{spinor}}=\boldsymbol\alpha_{SO(16)}+\boldsymbol\alpha'_{\mathrm{spinor}}$, or non-singular if otherwise. The $E_8$ root vectors $\boldsymbol\alpha_{E_8}$ now consists of the 112 $\boldsymbol\alpha_{SO(16)}$'s and 128 $\boldsymbol\alpha_{\mathrm{spinor}}$'s.

The $240=112+128$ $E_8$ roots can be generated by the 8 simple roots \begin{gather}\begin{pmatrix}|&\ldots&|\\\boldsymbol\alpha_1&\ldots&\boldsymbol\alpha_8\\|&\ldots&|\end{pmatrix}=\left(\begin{smallmatrix}
	0 & 0 & 0 & 0 & 0 & 0 & 0 & 1 \\
	1 & 0 & 0 & 0 & 0 & 0 & 0 & 0 \\
 -1 & 1 & 0 & 0 & 0 & 0 & 0 & 0 \\
	0 & -1 & 1 & 0 & 0 & 0 & 0 & 0 \\
	0 & 0 & -1 & 1 & 0 & 0 & 0 & 0 \\
	0 & 0 & 0 & -1 & 1 & 0 & 0 & 0 \\
	0 & 0 & 0 & 0 & -1 & 1 & 0 & 1 \\
	0 & 0 & 0 & 0 & 0 & -1 & 1 & 1 \\
	0 & 0 & 0 & 0 & 0 & 0 & -1 & 1 \\
 0 & 0 & 0 & 0 & 0 & 0 & 0 & 0 \end{smallmatrix}\right).\label{E8simpleroots}\end{gather} Their scalar products $\boldsymbol\alpha_I\cdot\eta\boldsymbol\alpha_J=(K_{E_8})_{IJ}$ recover the Cartan matrix of $E_8$ defined in Eq. \eqref{KE8}.

The embedding into the Minkowski space $\mathbb{R}^{1,9}$ guarantees a counter-propagating pair of redundant modes. They are the Dirac fermions $f_R\sim e^{i\boldsymbol\alpha_f^R\cdot\boldsymbol\phi}$ and $f_L\sim e^{i\boldsymbol\alpha_f^L\cdot\boldsymbol\phi}$, where $\boldsymbol\alpha^R_f={\bf e}_9$ and $\boldsymbol\alpha^L_f=3{\bf e}_0+{\bf e}_1+\ldots+{\bf e}_8$. They are fermionic because of the unit norm squares $\boldsymbol\alpha^R_f\cdot\eta\boldsymbol\alpha^R_f=1$ and $\boldsymbol\alpha^L_f\cdot\eta\boldsymbol\alpha^L_f=-1$. They decoupled from each other as well as the $E_8$ roots since $\boldsymbol\alpha_{E_8}\cdot\eta\boldsymbol\alpha^{R/L}_f=\boldsymbol\alpha^R_f\cdot\eta\boldsymbol\alpha^L_f=0$. Grouping $\boldsymbol\alpha^{R/L}_f$ with the $E_8$ simple roots in Eq. \eqref{E8simpleroots}, the $10\times10$ matrix $A=(\boldsymbol\alpha_1,\ldots,\boldsymbol\alpha_8,\boldsymbol\alpha^R_f,\boldsymbol\alpha^L_f)$ is unimodular (i.e.~$|\det(A)|=1$), and may be decomposed into block diagonal form as \begin{align}A^T\eta A=\begin{pmatrix}K_{E_8}&\\&\sigma_z\end{pmatrix},\label{AAT}\end{align} where $\sigma_z=\mathrm{diag}(1,-1)$.

Having completing the formal definition of the $E_8$ algebra, we now construct the gapping potential, which consists of inter-vortex $E_8$ current backscattering and intra-vortex fermion backscattering $f_Rf_L^\dagger$. Each vortex has chirality $c=16$ and carries 16 chiral Dirac fermions. We extend by vortex reconstruction to 18 forward moving Dirac fermions plus 2 backward moving ones. They can be bipartitioned into two groups of $9+1$, each of which can be transformed unimodularly into $E_8\times U(1)^R\times U(1)^L$, as has been explained above. In a similar fashion to our the previous discussions, we label the two groups by $A$ and $B$. The gapping potential is \begin{align}H&=-\frac{u_{\mathrm{inter}}}{2}\sum_{xy}\sum_{\boldsymbol\alpha_{E_8}}E^{A,\boldsymbol\alpha_{E_8}}_{x,y}E^{B,\boldsymbol\alpha_{E_8}}_{x,y+1}\nonumber\\&\;\;\;-\frac{u_{\mathrm{intra}}}{2}\sum_{xy}\sum_{C=A,B}(f_R^Cf_L^C+h.c.)\nonumber\\&=-u_{\mathrm{inter}}\sum_{xy}\sum_{\boldsymbol\alpha_{E_8}}\cos\left[\boldsymbol\alpha_{E_8}\cdot(\boldsymbol\phi^A_{x,y}+\boldsymbol\phi^B_{x,y+1})\right]\nonumber\\&\;\;\;-u_{\mathrm{intra}}\sum_{xy}\sum_{C=A,B}\cos\left(\boldsymbol\alpha^R_f\cdot\boldsymbol\phi^C_{xy}+\boldsymbol\alpha^L_f\cdot\boldsymbol\phi^C_{xy}\right)\label{E8sinegordon}\end{align} where the first sum runs over all 240 $E_8$ roots $\boldsymbol\alpha_{E_8}$.

All terms preserve the symmetries that we have defined in Eq. \eqref{bosonsymmtrans}. The angle order parameters $\Theta^{\boldsymbol\alpha_{E_8}}_{x,y+1/2}=\boldsymbol\alpha_{E_8}\cdot(\boldsymbol\phi^A_{x,y}+\boldsymbol\phi^B_{x,y+1})$ that appear in the inter-vortex term obey the symmetry relations \begin{align}T_A\Theta^{\boldsymbol\alpha_{E_8}}_{x,y+1/2}(z)T_A^{-1}&=-\Theta^{\boldsymbol\alpha_{E_8}}_{x,y+3/2}(z)+\pi\boldsymbol\alpha_{E_8}\cdot{\bf t}\nonumber\\G\Theta^{\boldsymbol\alpha_{E_8}}_{x,y+1/2}(z)G^{-1}&=-\Theta^{\boldsymbol\alpha_{E_8}}_{x+1,y+1/2}(-z)+\pi\boldsymbol\alpha_{E_8}\cdot{\bf t}\nonumber\\\mathsf{t}_{11}\Theta^{\boldsymbol\alpha_{E_8}}_{x,y+1/2}(z)\mathsf{t}_{11}^{-1}&=\Theta^{\boldsymbol\alpha_{E_8}}_{x+1,y+3/2}(z)\nonumber\\\mathsf{t}_{\bar{1}1}\Theta^{\boldsymbol\alpha_{E_8}}_{x,y+1/2}(z)\mathsf{t}_{\bar{1}1}^{-1}&=\Theta^{\boldsymbol\alpha_{E_8}}_{x-1,y+3/2}(z),\end{align} where ${\bf t}=(1,1,\ldots,1)^T$. The intra-vortex ones $\Theta^C_{x,y}=(\boldsymbol\alpha^R_f+\boldsymbol\alpha^L_f)\cdot\boldsymbol\phi^C_{xy}$, for $C=A,B$, obey \begin{align}T_A\Theta^C_{x,y}(z)T_A^{-1}&=-\Theta^C_{x,y+1}(z)+\frac{1-(-1)^{x+y}}{2}\pi(\boldsymbol\alpha^R_f+\boldsymbol\alpha^L_f)\cdot{\bf t}\nonumber\\G\Theta^C_{x,y}(z)G^{-1}&=-\Theta^C_{x+1,y}(-z)+\frac{\pi}{2}(\boldsymbol\alpha^R_f+\boldsymbol\alpha^L_f)\cdot{\bf t}\nonumber\\\mathsf{t}_{11}\Theta^C_{x,y}(z)\mathsf{t}_{11}^{-1}&=\Theta^C_{x+1,y+1}(z)\nonumber\\\mathsf{t}_{\bar{1}1}\Theta^C_{x,y}(z)\mathsf{t}_{\bar{1}1}^{-1}&=\Theta^C_{x-1,y+1}(z).\end{align} For the $E_8$ order parameters, $\boldsymbol\alpha_{E_8}\cdot{\bf t}=0,\pm4,\pm8,\pm12$ and, therefore, the inter-vortex sine-Gordon terms in Eq. \eqref{bosonsymmtrans} are symmetric, since $\cos(-\Theta+2m\pi)=\cos\Theta$. The intra-vortex terms in Eq. \eqref{bosonsymmtrans} are also symmetric because $(\pi/2)(\boldsymbol\alpha^R_f+\boldsymbol\alpha^L_f)\cdot{\bf t}=6\pi$. The ground state expectation values $\left\langle\Theta^{\boldsymbol\alpha_{E_8}}_{x,y+1/2}\right\rangle$ transform consistently according to the symmetries (c.f.~Eq.\eqref{evenorderparameter} for the even $N$ case). The fermionic order parameters $\left\langle\Theta^C_{x,y}\right\rangle$ transform consistently up to large gauge transformation $\left\langle\Theta^C_{x,y}\right\rangle\equiv\left\langle\Theta^C_{x,y}\right\rangle+2\pi$ originated from the gauge redundancy $d^a\sim e^{i\phi^a}=e^{i(\phi^a+2\pi)}$.

\begin{figure}[htbp]
\centering\includegraphics[width=0.4\textwidth]{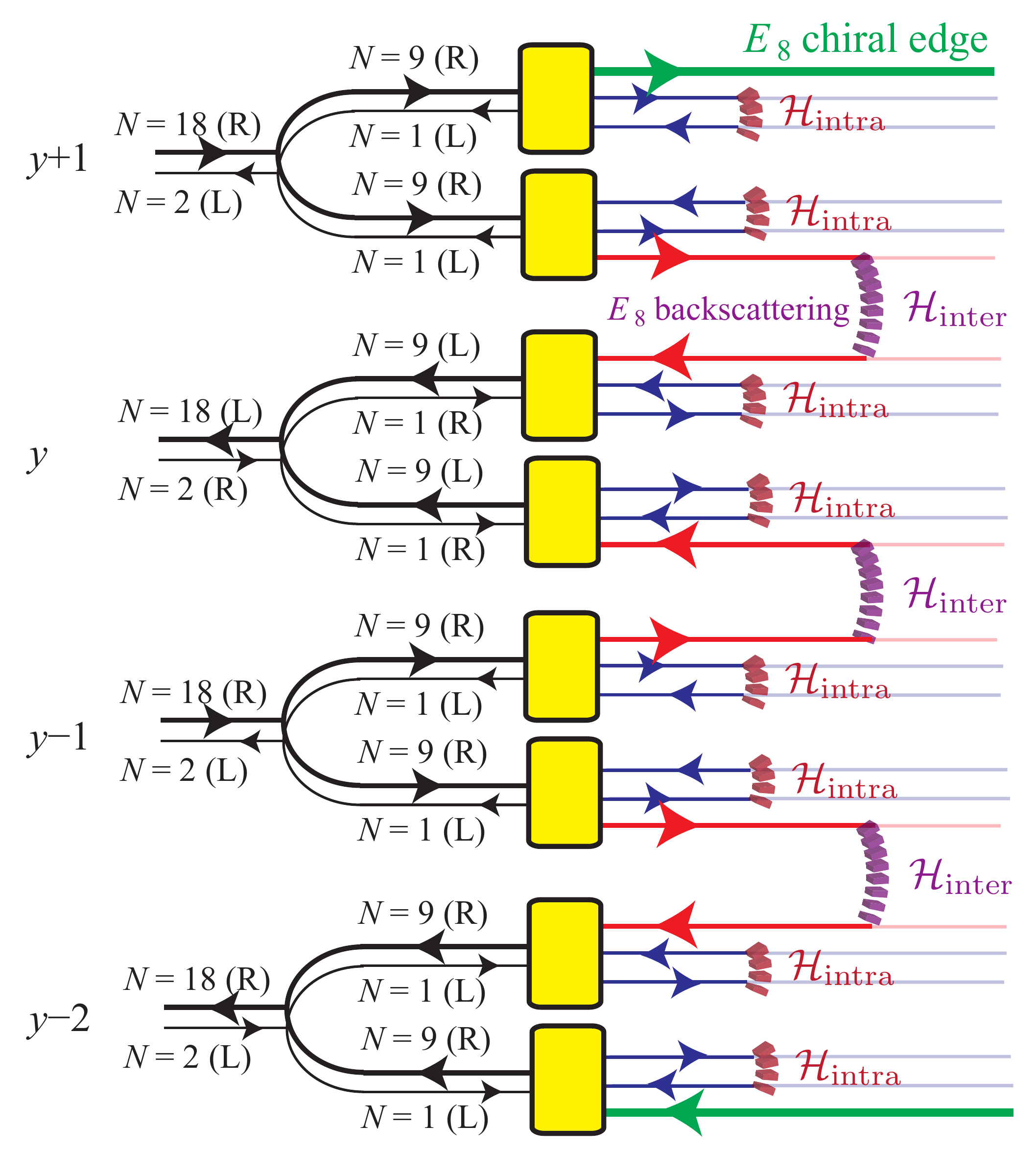}
\caption{Schematic figure of $E_8$ decomposition and gapping interaction. Dirac modes (black lines) along each vortex $y$ are decomposed into two sets of $E_8$ chiral CFT (red lines) and two counter-propagating pairs of Dirac fermions (blue lines) by the basis transformation $A$ in Eq. \eqref{AAT} (yellow boxes). The $E_8$ backscattering terms $\mathcal{H}_{\mathrm{inter}}$ and the fermion backscattering $\mathcal{H}_{\mathrm{intra}}$ are defined in Eq. \eqref{E8sinegordon}. Uncoupled $E_8$ chiral CFTs are left along the boundaries of an open system, while the bulk mimics the $E_8$ quantum Hall state.}\label{fig:E8schematic}
\end{figure}

The angle variables in Eq. \eqref{E8sinegordon} satisfy the Haldane nullity condition, and the sine-Gordon interaction generates a symmetry preserving gap in the energy spectrum. In Eq. \eqref{E8sinegordon}, we considered the $E_8$ backscattering term, coupling with adjacent wires in $\hat{y}$-direction. In the absence of the single-body hopping but Eq. \eqref{E8sinegordon}, the arrays of the Dirac strings form a coupled $2D$ layer in $yz$ plane, possessing a finite bulk gap. We now consider open boundary condition in $y$ direction as shown in Fig.~\ref{fig:E8schematic}. Along the boundary of the $2D$ layer, a chiral $E_8$ edge state is left uncoupled and remains gapless, since there is no counter-propagating adjacent Dirac string to pair with. Therefore, each $2D$ layer resembles a quantum Hall state carrying a $E_8$ CFT as its edge theory in low-energy. As a result, the $2D$ bulk topological order and the bulk excitations of the layers can be inferred from the $E_8$ edge state.

The low energy effective theory for the $E_8$ quantum Hall states are described by Chern-Simons theory with $K_{E_8}$ matrix whose action is given as\cite{PhysRevB.86.125119},
\begin{gather}
\mathcal{S}_{cs}=\frac{1}{4\pi}\sum_{I,J}\int dx^3  (K_{E_8})_{IJ} \epsilon^{\mu\nu\lambda} a_{I,\mu}\partial_{\nu}a_{J,\lambda}-\sum_{I}a_{\mu,I}j_{\mu, I}
\end{gather}
and $K_{E_8}$ is the Cartan matrix of $E_8$ defined in Eq. \eqref{KE8}. Here, $a_I$ is $I^{\mathrm{th}}$ dynamical Abelian gauge field. $K_{E_8}$ matrix contains the information of the bulk quasi-particle excitations and the corresponding edge theory. To be specific, the topological order or the ground state degeneracy of the corresponding bulk theory is identified as the determinant of $K$ matrix. It can be seen that the bulk topological order of $E_8$ state is trivial ($\det(E_8)=1$), since it is unimodular. Consequently, the system only supports local exitations with non-fractional statistics.

We have constructed the symmetry preserving many-body gapped phase with no topological order. It is important to note that the presence of such a phase is intimately related to the properties of even unimodular lattice. 
The $E_8$ lattice is the minimal even unimodular lattice, that appears in $8$ dimensions. Other even unimodular lattices, such as the Leech lattice in dimension 24, that appear in higher dimensions can be similarly utilized to construct the topologically trivial gapping potential.

\section{Cancellation of the Large Gravitational Anomaly}\label{secmodreview}
In the previous section, we constructed the gapping potential when $N=16=8+8$ fermion flavors are present by utilizing the $E_8 \times E_8$ decomposition. By examining the topological properties of the Chern-Simons theory, characterized by the Cartan matrix, $K_{E_8}$, we show that the lack of topological order is directly connected to the fact that the root lattice of $K_{E_8}$ matrix forms the even unimodular lattice. We additionally show that the presence of the unimodular $E_8$ lattice in $8$ dimensions is intimately related to the cancellation of the large gravitational anomaly that is known to exist in topologically trivial systems.

To begin our discussion, we consider the situation where the vortex line of the Dirac string forms a periodic ring of circumference, $\it{l}$ at the inverse temperature, $\beta$. Then, the Dirac string lives on a $(1+1)$-D compact space-time manifold $T^1\times T^1$ where $T^1$ is a torus. The $T^1\times T^1$ space-time manifold can be described by the modular parameter $\tau=\it{l}_0/\it{l}_1$, where $\it{l}_0$ and $\it{l}_1$ are the periods of the space and Euclidean time coordinates, respectively. On the torus characterized by the modular parameter $\tau$, the classical action of the right-moving Dirac string can be written as,
\bea
\label{Eq:1Daction}
S=-\int_0^1dz\int_0^1d\tau\bar R_{\lambda,\mu}(\it{l}_1\partial_\tau+\it{l}_0\partial_z)R_{\lambda,\mu}
\eea
$\lambda$ and $\mu$ specifies the boundary condition on the space and time respectively such that the chiral fermion follows,
\bea
R_{\lambda,\mu}(\tau+\beta,z)=e^{2\pi i \mu}R_{\lambda,\mu}(\tau,z),
\\
\nonumber
R_{\lambda,\mu}(\tau,z+\it{l})=e^{2\pi i \lambda}R_{\lambda,\mu}(\tau,z).
\eea
For example, $\lambda=0(1/2)$ indicates the presence of periodic (anti-periodic boundary conditions). The partition function of the single right-moving Dirac string can be evaluated as\cite{bigyellowbook,PhysRevB.90.165134},
\begin{gather}
Z_{\mu,\lambda}
=\prod_{n=-\infty}^\infty (1-e^{2\pi i \mu}e^{2\pi i \tau (n+\lambda)})
\nonumber
\\
=
\frac{1}{\eta(\tau)}\theta[\lambda-\frac{1}{2},\mu-\frac{1}{2}](\tau),
\label{Eq1Dpartition}
\end{gather}
where $\tau$ is the modular parameter and $H$ is the Hamiltonian for the given theory. Further, in Eq. \eqref{Eq1Dpartition} $\theta$ and $\eta$ are the Jacobi theta function and the Dedekind eta function respectively. For completeness, these functions are explicitly given as,
\begin{gather}
\label{Eq:theta}
\eta(\tau)=q^{1/24}\prod_{n=1}^{\infty} (1-q^n),
\\
\nonumber
\theta[\lambda,\mu](\tau)=\sum_{n=-\infty}^{n=\infty} e^{2\pi i (n+\lambda) \mu} q^{(n+\lambda)^2/2} .
\end{gather}
where $q=e^{2\pi i \tau}$. $T^1\times T^1$ possesses a set of discrete coordinate transformations that effectively map the toroidal space back onto itself, and are referred to as the modular transformation. More specifically, the modular transformation can be defined as a group of transformations on the modular parameter, $\tau$, that are given as
\bea
\label{Eq:Mod}
\tau \rightarrow \frac{a\tau+b}{c\tau+d},
\eea
where $a,b,c,d$ are integers, which satisfying $ad-bc=1$. There exist two generators that form the modular group: $S$ and $T$, which transforms the modular parameter as $\tau\rightarrow -\frac{1}{\tau}$ and $\tau\rightarrow \tau+1$ respectively. The action in Eq. \eqref{Eq:1Daction} is invariant under the modular transformations. Therefore, it is a classical symmetry that the Dirac strings possess.

However, the partition function in Eq. \eqref{Eq1Dpartition} is not invariant under the modular transformation. Under $S$ and $T$ transformations, the partition function of the Dirac string explicitly transforms accordingly,
\begin{gather}
\nonumber
T[Z_{R,\lambda,\mu}]=e^{-\pi i (\lambda^2-\lambda+1/6)}Z_{R,\lambda,\mu+\lambda},
\\
S[Z_{R,\lambda,\mu}]=e^{-2\pi i (\lambda-1/2)(\mu-1/2)}Z_{R,\mu,-\lambda}.
\label{Eq:modularphase}
\end{gather}
where the partition function of the right-moving fermion, $Z_L$, gains the opposite phase. Indeed, the partition function is not invariant on the quantum level. Therefore, the single $1D$ chiral fermion possesses the large gravitational anomaly under the modular transformations\cite{PhysRevB.85.245132}.

After the basic calculation of the gravitational anomaly of the single Dirac string, we are now interested in the situation where the product of the left and the right moving Dirac strings cancels out the anomaly, which indicates that the theory is trivial without topological order. We seek a particular combination of the partition functions such that $Z_{total}=\sum_{\lambda_L,\lambda,\mu_L,\mu_R}Z_{L,\lambda_L,\mu_L}Z_{R,\lambda_R,\mu_R}$. For example, without any symmetry constraints, there is an obvious combination that is given by,
\begin{gather}
\nonumber
Z_{total}=Z_{L,0,0}Z_{R,0,0}+Z_{L,\frac{1}{2},0}Z_{R,\frac{1}{2},0}
\\
+Z_{L,0,\frac{1}{2}}Z_{R,0,\frac{1}{2}}+Z_{L,\frac{1}{2},\frac{1}{2}}Z_{R,\frac{1}{2},\frac{1}{2}}.
\end{gather}
The above partition function is invariant under the $S$ and the $T$ transformations.

However, we are interested in solving for the partition function that represent the superconducting media. In this case, the system intrinsically possesses the fermion number parity symmetry. To examine the anomaly in our model, we force the fermion parity symmetry in our partition function. We wish to check if the partition function of the sub-Hilbert space, possessing a definite fermion parity, is modular invariant. To do so, we project the total Hilbert space into the subsector by considering the projection operator, which is given as,
\bea
P_{L(R)}=\frac{1+(-1)^{n_{L(R)}}}{2}
\eea
where $n_{L(R)}$ is the fermion number operator. The above projection operator is $1$ if the fermion number is even, and zero if it is odd. The projection operator maps into the even fermion number sector of the Hilbert space. By inserting the projection operator to the partition function, the right-moving partition function with a definite fermion parity can be evaluated as,
\begin{gather}
Z_{R,\lambda}=\mathrm{Tr}(P_R  e^{2\pi i \tau H})
\\
\nonumber
=\frac{1}{2}\mathrm{Tr}((1+(-1)^{n_{R}}) e^{ 2 \pi i \tau H_R})
\\
\nonumber
=\frac{1}{2}\sum_{\mu=0,1/2}\mathrm{Tr}(e^{2\pi i \mu \sum_{a=1}^N n_R^a} e^{2\pi i \tau \sum_{a=1}^N H_R^a})
\\
\nonumber
=\frac{1}{2}(Z_{\mu=0,\lambda}^N+Z_{\mu=1/2,\lambda}^N).
\end{gather}
 Among the two generators, $S$ and $T$, of the modular transformation, we first demand that the partition function is invariant under $S$ transformation. $S$ invariant partition function must have the form of,
\begin{gather}
Z_{R}=Z_{R,\lambda=0}+Z_{R,\lambda=1/2}
\\
\nonumber
=\frac{1}{2}(Z_{0,0}^N+Z_{1/2,0}^N+Z_{0,1/2}^N+Z_{1/2,1/2}^N)
\\
\nonumber
=\frac{(\theta[0,0]^N+\theta[1/2,0]^N+\theta[0,1/2]^N+\theta[1/2,1/2]^N)(\tau)}{2\eta(\tau)^N}
\end{gather}
We now find that the partition function having a definite fermion parity is given as a linear combinations of different boundary sectors.

If the different boundary sectors of $Z_R$ gains a single covariant phase under the modular transformations, $Z_{total}$ becomes the modular invariant. According to Eq. \eqref{Eq:modularphase}, each theta function of the different boundary sectors gains different phases under the $T$ transformation. Therefore, $Z_R$ is not modular covariant. However, when $N=8$ copies of the Dirac string exist, the above sum of the Jacobi theta function can be written as the theta function of $E_8$ lattice. Therefore, the partition function can be re-expressed as,
\bea
Z_R=\frac{1}{2\eta(\tau)^N}\sum_{x\in \Lambda_{E_8}} e^{\pi i\tau|x|^2}
=\frac{1}{2\eta(\tau)^N}\Theta(\tau)_{E_8}
\label{Eq:thetae8}
\eea
where $\Lambda_{E8}$ is the set of the $E_8$ lattice. The theta function of the $E_8$ lattice is the modular form of weight $8$, which is invariant under the $T$ transformation. Furthermore, the theta function in Eq. \eqref{Eq:thetae8} can be explicitly transformed into the Minkowski space embedded $E_8$ lattice which we used in Eq. \eqref{E8simpleroots}.
\bea
\Theta(\tau)_{E_8,M}\equiv\sum_{ \vec{x}\in E_{8,M}} e^{\pi i\tau x\cdot \eta x}=\Theta(\tau)_{E_8},
\eea
where we relegate the explicit form of the transformations to Appendix \ref{Appendix:E8}. As a result, we find that the modular covariance property of $Z_R$ can be captured by the transformation between the partition function and the theta function of $E_8$.

In addition to the contribution of the theta function, the $\eta$ function on the denominator gains a phase $e^{\frac{8\pi i}{12}}$ under the $T$ transformation, because the Dedekind eta function satisfies the following property:
\bea
\eta(\tau+1)=e^{\pi i /12}\eta(\tau),\quad
\eta(-1/\tau)=\sqrt{-i\tau}\eta(\tau)
\label{Eq:eta}
\eea
 Therefore, $Z_{R}$ satisfies the modular covariance under the $S$ and $T$ transformations as,
\bea
Z_{R}(\tau)=Z_R(-\frac{1}{\tau}),\quad Z_{R}(\tau)=e^{\frac{8\pi i}{12}}Z_R(\tau+1)
\eea
Therefore the total partition function, $Z_{total}=Z_{L}Z_{R}=|Z_{L}|^2$, which consists of $8$ left-moving and $8$ right-moving Dirac strings, establishes the modular invariance.

Before we conclude this section, we would like to comment on the possible implication of the modular invariance on the interacting topological classification. The modular covariance when $N=16$ should not be confused with $\mathbb{Z}_{16}$ classification of 3D DIII time-reversal symmetric topological superconductor. In the coupled wire setting, $T_A$ is no longer local time-reversal symmetry as it is accompanied by the additional translation symmetry. Since the symmetry classes are different, we cannot exactly compare our model with the DIII class. For example, this difference is already seen in section \ref{sec:coupled} within the single-body Hamiltonian, showing that our model already has $\mathbb{Z}_2$ classification, which is different from the $\mathbb{Z}$ classification of 3D DIII class. Nevertheless, our analysis in Eq. \eqref{Eq:thetae8} shows that the modular covariance is attributed to the presence of $E_8$ lattice. We thus conclude that the modular covariance coincides with the topologically trivial decomposition $SO(32) \sim E_8 \times E_8$ when $N=16=8+8$. In addition to the modular covariance at $N=8$, $Z_R$ can achieve the modular invariance if the phase of the eta function under $\tau \rightarrow \tau+1$ transformation in Eq. \eqref{Eq:eta} cancels out. This leads to the modular invariance of the chiral partition function, $Z_{R}$, when $N=24$, as it has been pointed out in previous work\cite{PhysRevB.85.245132}.
\section{Conclusion}\label{seccon}
In this work, we have explicitly constructed various forms of many body interactions that open gaps in the energy spectrum while preserving the underlying symmetries present in coupled wire constructions of $3D$ Dirac nodal superconductors. In section \ref{sec:manybody1}, we found that the gapped bulk of the two-dimensional $y-z$ plane supports non-local fractional quasi-particles that develop the non-trivial topological orders. When the system is extended into the full three-dimensions, the fractional excitations can be still maintained to generate the topological degeneracy. In this work, we indicate that the many-body interactions generate non-trivial topological orders in three dimensions. However, it still remains unresolved that how these non-local excitations behave in three dimensions. The detailed physical behavior of these fractional particles can be studied in future works.

Furthermore, we constructed a unimodular $E_8$ gapping potential when there are $N=16$ Dirac channels along a vortex line. To build the $E_8$ gapping potential, we utilized $SO(32)\sim E_8 \times E_8$ decomposition. The resulting gapped phase did not support the topological order due to the unimodular property of the $E_8$ lattice. In general, even unimodular lattices exist in every dimensions multiples of $8$. For instance, in $16$ dimensions, two even unimodular lattice exist. We can combine the two $E_8$ lattices to build $E_8\oplus E_8$, or we can consider the $D_{16+}$ lattice that tightly packs spheres in the $16$ dimensional checker board lattice as $E_8$ similarly does in the $8$ dimensions. More interestingly, in $24$ dimensions, the Leech lattice emerges. The Leech lattice is the simplest even unimodular lattice without any root vectors. This property of the Leech lattice is related to the emergence of the perfect $1D$ metal phase that are immune to localizations\cite{PhysRevB.90.241101}. To this end, it would be interesting to study the properties of the different unimodular gapping potentials in the higher dimensions.

Finally, we have shown that the presence of the $E_8$ gapping potential is reflected by the cancellation of the large gravitational anomaly. In gapped phases, the quantum anomalies can be understood as the physical response of the edge states. For instance, the gravitational anomaly can be interpreted as the thermal pumping\cite{PhysRevB.95.165405} in the quantum Hall systems. We can perform the similar analysis in the semimetallic or nodal superconductor phases by supporting auxiliary higher dimensional bulk theory. For example, we can stack the 3D Weyl semimetallic phases to construct the 4D quantum Hall phase. When the nodal points are physically separated by the 4D bulk, we can similarly consider the pumping argument between the 3D surfaces. It would be interesting to study how these anomalies manifest as physical responses in the interacting systems.

\acknowledgments
SR and JCYT are supported by the National Science Foundation under Grant No.~DMR-1653535. MJP and MJG are supported by National Science Foundation under grant No. DMR 17-10437. MJG acknowledges financial support from the Office of Naval Research (ONR) under grant number N00014-17-1-3012.

\bibliography{reference}

\onecolumngrid
\appendix
\section{Symmetry transformation of chiral vortex}\label{Appendix:A}
In this appendix, we verify the symmetry actions in defined Eq. (\ref{CWsymmetries}) by explicitly solving for the chiral vortex states in the continuum model. In the Hamiltonian in Eq. \eqref{varray}, each Weyl species labeled by $\mu$ degree of the freedom is decoupled from each other. We can solve them separately. The Hamiltonian of $\mu=1$ sector is explicitly written as,
\begin{gather}
H_{\mathrm{SC-Dirac}}({\bf k},{\bf r})=\left(
\begin{array}{cccc}
 k_z &k_- & |\Delta| e^{-i\phi} & 0  \\
k_+ & -k_z & 0  & |\Delta| e^{-i\phi} \\
|\Delta| e^{i\phi} & 0 & -k_z & -k_-\\
0  & |\Delta| e^{i\phi} & -k_+ & k_z \\
\end{array}
\right)
\end{gather}
where we omit the factor, $\hbar v_f$. We set $\Delta(\vec{r})=|\Delta|e^{-\phi}$
We rewrite the Hamiltonian in real space and approximate $k_z=0$. Then, the Hamiltonian becomes
\begin{gather}
H_{\mathrm{SC-Dirac}}({\bf k},{\bf r})=
\left(
\begin{array}{cccc}
 0 &-ie^{-i\phi}(\partial_r-\frac{i}{r}\partial_\phi) & i|\Delta| e^{-i\phi} & 0  \\
ie^{i\phi}(\partial_r+\frac{i}{r}\partial_\phi) & 0 & 0  & i|\Delta| e^{-i\phi} \\
-i|\Delta| e^{i\phi} & 0 & 0 & ie^{-i\phi}(\partial_r-\frac{i}{r}\partial_\phi)\\
0  & -i|\Delta| e^{i\phi} & -ie^{i\phi}(\partial_r+\frac{i}{r}\partial_\phi) & 0 \\
\end{array}
\right)
\end{gather}
The above Hamiltonian possesses a zero mode vortex solution, which is given as,
\bea
\frac{1}{r}e^{-|\Delta| r}(-ie^{-i\phi},0,0,i e^{i\phi})^T
\eea
The above solution has the dispersion $k_z$ when $k_z\neq 0$. In the similar manner, when we translate $\vec{r}\rightarrow \vec{r}+\vec{e}_{x,y}$, $|\Delta|$ transforms to $-|\Delta|$. The vortex solution can be similarly obtained as,
\bea
\frac{1}{r}e^{-|\Delta| r}(-ie^{-i\phi},0,0,i e^{i\phi})^T
\eea
When we translate $\vec{r}\rightarrow \vec{r}+\frac{\vec{e}_{x}\pm \vec{e}_{y}}{2}$, $|\Delta|e^{i\phi}$ transforms to $\pm |\Delta|e^{-\phi}$. In this case, the zero modes have a dispersion given as, $-k_z$. When $\Delta(\vec{r})=|\Delta|e^{-i\phi}$, the vortex solution can be obtained as,
\bea
\frac{1}{r}e^{-|\Delta| r}(0,ie^{i\phi},i e^{-i\phi},0)^T
\eea
When $\Delta(\vec{r})=-|\Delta|e^{-i\phi}$, the vortex solution can be obtained as,
\bea
\frac{1}{r}e^{-|\Delta| r}(0,ie^{i\phi},-i e^{-i\phi},0)^T
\eea

We similarly solve for $\mu_z=-1$ sector to write down the full solution of $8\times 8$ Hamiltonian. In the basis of $\eta=(c^\dagger_{\uparrow,\mu=1},c^\dagger_{\downarrow,\mu=1},c_{\downarrow,\mu=1},-c_{\uparrow,\mu=1},c^\dagger_{\uparrow,\mu=-1},c^\dagger_{\downarrow,\mu=-1},c_{\downarrow,\mu=-1},-c_{\uparrow,\mu=-1})^T$,
the solutions below define the Dirac fermion operators $L$ and $R$. For $\Delta(\vec{r})=|\Delta|e^{i\phi}$,
\bea
R_{x,y}=\frac{1}{r}e^{-|\Delta| r}(-ie^{-i\phi},0,0, ie^{i\phi},0,ie^{i\phi},-ie^{-i\phi},0) \eta.
\eea
For $\Delta(\vec{r}+\vec{e}_{x,y})=-|\Delta|e^{i\phi}$,
\bea
R_{x+1,y+1}=R_{x-1,y+1}^\dagger=\frac{1}{r}e^{-|\Delta| r}(ie^{i\phi},0,0,i e^{-i\phi},0,ie^{i\phi},i e^{-i\phi},0) \eta.
\eea
For $\Delta(\vec{r}+\frac{\vec{e}_{x}+ \vec{e}_{y}}{2})=|\Delta|e^{-i\phi}$,
\bea
L_{x,y+1}=\frac{1}{r}e^{-|\Delta| r}
(0,ie^{i\phi},i e^{-i\phi},0,ie^{-i\phi},0,0,i e^{i\phi}) \eta.
\eea
For $\Delta(\vec{r}+\frac{\vec{e}_{x}- \vec{e}_{y}}{2})=-|\Delta|e^{-i\phi}$,
\bea
L_{x+1,y}=\frac{1}{r}e^{-|\Delta| r}
(0,-ie^{i\phi},i e^{-i\phi},0,ie^{-i\phi},0,0,-i e^{i\phi}) \eta.
\eea
The above solutions of the Dirac operators can be checked to satisfy the set of the symmetry transformation rules defined in Eq. (\ref{CWsymmetries}), according to the symmetry operators defined in the continuum limit in Eq. \eqref{DiracNODALHTR}, \eqref{DiracNODALHglide}. To see this, we explicitly list the symmetry operations of the vortex operators. We omit the $\frac{1}{r}e^{-|\Delta| r}$ and $\eta$ for the simplicity. Under the lattice translational $t_x$ and $t_y$ symmetries, the Dirac strings transform as,
\bea
\mu_y \tau_y R_{x,y}=\left(
\begin{array}{cccccccc}
 0 & 0 & 0 & 0 & 0 & 0 & -1 & 0 \\
 0 & 0 & 0 & 0 & 0 & 0 & 0 & -1 \\
 0 & 0 & 0 & 0 & 1 & 0 & 0 & 0 \\
 0 & 0 & 0 & 0 & 0 & 1 & 0 & 0 \\
 0 & 0 & 1 & 0 & 0 & 0 & 0 & 0 \\
 0 & 0 & 0 & 1 & 0 & 0 & 0 & 0 \\
 -1 & 0 & 0 & 0 & 0 & 0 & 0 & 0 \\
 0 & -1 & 0 & 0 & 0 & 0 & 0 & 0 \\
\end{array}
\right)
\left(
\begin{array}{c}
 -i e^{-i \phi } \\
 0 \\
 0 \\
 i e^{i \phi } \\
 0 \\
 i e^{i \phi } \\
 -i e^{-i \phi } \\
 0 \\
\end{array}
\right)=
\left(
\begin{array}{c}
 i e^{-i \phi } \\
 0 \\
 0 \\
 i e^{i \phi } \\
 0 \\
 i e^{i \phi } \\
 i e^{-i \phi } \\
 0 \\
\end{array}
\right)=R_{x+1,y+1}
\eea
\bea
\mu_y\tau_y L_{x,y+1}=\left(
\begin{array}{cccccccc}
 0 & 0 & 0 & 0 & 0 & 0 & -1 & 0 \\
 0 & 0 & 0 & 0 & 0 & 0 & 0 & -1 \\
 0 & 0 & 0 & 0 & 1 & 0 & 0 & 0 \\
 0 & 0 & 0 & 0 & 0 & 1 & 0 & 0 \\
 0 & 0 & 1 & 0 & 0 & 0 & 0 & 0 \\
 0 & 0 & 0 & 1 & 0 & 0 & 0 & 0 \\
 -1 & 0 & 0 & 0 & 0 & 0 & 0 & 0 \\
 0 & -1 & 0 & 0 & 0 & 0 & 0 & 0 \\
\end{array}
\right)\left(
\begin{array}{c}
 0 \\
 i e^{i \phi } \\
 i e^{-i \phi } \\
 0 \\
 i e^{-i \phi } \\
 0 \\
 0 \\
 i e^{i \phi } \\
\end{array}
\right)=\left(
\begin{array}{c}
 0 \\
 -i e^{i \phi } \\
 i e^{-i \phi } \\
 0 \\
 i e^{-i \phi } \\
 0 \\
 0 \\
 -i e^{i \phi } \\
\end{array}
\right)=L_{x+1,y}
\eea
Under the glide mirror symmetry, the Dirac strings transform as,
\bea
GR_{x,y}=s_z\tau_y R_{x,y}=
\left(
\begin{array}{cccccccc}
 0 & 0 & -i & 0 & 0 & 0 & 0 & 0 \\
 0 & 0 & 0 & i & 0 & 0 & 0 & 0 \\
 i & 0 & 0 & 0 & 0 & 0 & 0 & 0 \\
 0 & -i & 0 & 0 & 0 & 0 & 0 & 0 \\
 0 & 0 & 0 & 0 & 0 & 0 & -i & 0 \\
 0 & 0 & 0 & 0 & 0 & 0 & 0 & i \\
 0 & 0 & 0 & 0 & i & 0 & 0 & 0 \\
 0 & 0 & 0 & 0 & 0 & -i & 0 & 0 \\
\end{array}
\right)\left(
\begin{array}{c}
 -i e^{-i \phi } \\
 0 \\
 0 \\
 i e^{i \phi } \\
 0 \\
 i e^{i \phi } \\
 -i e^{-i \phi } \\
 0 \\
\end{array}
\right)
=\left(
\begin{array}{c}
 0 \\
 e^{i \phi } \\
 -e^{-i \phi } \\
 0 \\
 e^{-i \phi } \\
 0 \\
 0 \\
 -e^{i \phi } \\
\end{array}
\right)
=i\left(
\begin{array}{c}
 0 \\
 -i e^{i \phi } \\
 i e^{-i \phi } \\
 0 \\
 -i e^{-i \phi } \\
 0 \\
 0 \\
 i e^{i \phi } \\
\end{array}
\right)=iL_{x+1,y}^\dagger
\eea

\bea
GL_{x,y+1}=s_z\tau_y L_{x,y+1}=\left(
\begin{array}{cccccccc}
 0 & 0 & -i & 0 & 0 & 0 & 0 & 0 \\
 0 & 0 & 0 & i & 0 & 0 & 0 & 0 \\
 i & 0 & 0 & 0 & 0 & 0 & 0 & 0 \\
 0 & -i & 0 & 0 & 0 & 0 & 0 & 0 \\
 0 & 0 & 0 & 0 & 0 & 0 & -i & 0 \\
 0 & 0 & 0 & 0 & 0 & 0 & 0 & i \\
 0 & 0 & 0 & 0 & i & 0 & 0 & 0 \\
 0 & 0 & 0 & 0 & 0 & -i & 0 & 0 \\
\end{array}
\right)\left(
\begin{array}{c}
 0 \\
 i e^{i \phi } \\
 i e^{-i \phi } \\
 0 \\
 i e^{-i \phi } \\
 0 \\
 0 \\
 i e^{i \phi } \\
\end{array}
\right)
=\left(
\begin{array}{c}
 -e^{-i \phi } \\
 0 \\
 0 \\
 -e^{i \phi } \\
 0 \\
 e^{i \phi } \\
 e^{-i \phi } \\
 0 \\
\end{array}
\right)=i\left(
\begin{array}{c}
 ie^{-i \phi } \\
 0 \\
 0 \\
 ie^{i \phi } \\
 0 \\
 -ie^{i \phi } \\
 -ie^{-i \phi } \\
 0 \\
\end{array}
\right)
=iR_{x+1,y+1}^\dagger
\eea
Finally, under the time-reversal symmetry, the Dirac strings transform as,
\bea
TR_{x,y}=is_y K R_{x,y}=\left(
\begin{array}{cccccccc}
 0 & 1 & 0 & 0 & 0 & 0 & 0 & 0 \\
 -1 & 0 & 0 & 0 & 0 & 0 & 0 & 0 \\
 0 & 0 & 0 & 1 & 0 & 0 & 0 & 0 \\
 0 & 0 & -1 & 0 & 0 & 0 & 0 & 0 \\
 0 & 0 & 0 & 0 & 0 & 1 & 0 & 0 \\
 0 & 0 & 0 & 0 & -1 & 0 & 0 & 0 \\
 0 & 0 & 0 & 0 & 0 & 0 & 0 & 1 \\
 0 & 0 & 0 & 0 & 0 & 0 & -1 & 0 \\
\end{array}
\right)\left(
\begin{array}{c}
 -i e^{-i \phi } \\
 0 \\
 0 \\
 i e^{i \phi } \\
 0 \\
 i e^{i \phi } \\
 -i e^{-i \phi } \\
 0 \\
\end{array}
\right)^*=\left(
\begin{array}{c}
 0 \\
 i e^{i \phi } \\
 i e^{-i \phi } \\
 0 \\
 i e^{-i \phi } \\
 0 \\
 0 \\
 i e^{i \phi } \\
\end{array}
\right)=L_{x,y+1}
\eea
\bea
TL_{x,y-1}=is_y K L_{x,y-1}
=\left(
\begin{array}{cccccccc}
 0 & 1 & 0 & 0 & 0 & 0 & 0 & 0 \\
 -1 & 0 & 0 & 0 & 0 & 0 & 0 & 0 \\
 0 & 0 & 0 & 1 & 0 & 0 & 0 & 0 \\
 0 & 0 & -1 & 0 & 0 & 0 & 0 & 0 \\
 0 & 0 & 0 & 0 & 0 & 1 & 0 & 0 \\
 0 & 0 & 0 & 0 & -1 & 0 & 0 & 0 \\
 0 & 0 & 0 & 0 & 0 & 0 & 0 & 1 \\
 0 & 0 & 0 & 0 & 0 & 0 & -1 & 0 \\
\end{array}
\right)\left(
\begin{array}{c}
 0 \\
 -i e^{-i \phi } \\
 -i e^{i \phi } \\
 0 \\
 -i e^{i \phi } \\
 0 \\
 0 \\
 -i e^{-i \phi } \\
\end{array}
\right)=\left(
\begin{array}{c}
 i e^{-i \phi } \\
 0 \\
 0 \\
 -i e^{i \phi } \\
 0 \\
 -i e^{i \phi } \\
 i e^{-i \phi } \\
 0 \\
\end{array}
\right)=-R_{x,y}
\eea
\section{Commutation relations of the bosonic operator}\label{Appendix:B}

In Eq. \eqref{ETCR}, we defined the equal-time commutation relation(ETCR) of the bosonic operator, which is given as,
\begin{gather}
\left[\phi_{x,y}^a(z),\phi_{x',y'}^{a'}(z')\right]
=c_{x,y,x',y'}^{a,a'}(z-z')
\nonumber
\\
i\pi(-1)^{\mathrm{min}\{x,x'\}+\mathrm{min}\{y,y'\}}\big[\delta_{xx'}\delta_{yy'}\delta_{aa'}\mathrm{sgn}(z'-z)
+\sigma_z\delta_{aa'}\delta_{yy'}\mathrm{sgn}(x-x')+\sigma_z\delta_{yy'}\mathrm{sgn}(a-a')-\sigma_z\mathrm{sgn}(y-y')\big].
\end{gather}
The alternating factor, $(-1)^{\mathrm{min}\{x,x'\}+\mathrm{min}\{y,y'\}}$, ensures that the adjacent wires counter-propagate each other while the ETCR is antisymmetric under the interchange of $(x,y,z,a)\Leftrightarrow (x',y',z',a')$. By taking the derivatives of $z$ on the ETCR, we find that the first term gives the canonical commutation relation between the conjugate bosonic fields, which is given as,
\bea
\left[\phi_{x,y}^a(z),\partial_{z'}\phi_{x',y'}^{a'}(z')\right]=
2\pi i(-1)^{x+y}\delta_{xx'}\delta_{yy'}\delta_{aa'}\delta(z'-z).
\eea
$(-1)^{\mathrm{min}\{x,x'\}+\mathrm{min}\{y,y'\}}$ is now replaced as $(-1)^{x+y}$, taking account into the change of the propagating direction.

The remaining terms in \eqref{ETCR} ensure the correct anticommutation relations $ \left\{ e^{\pm i \phi_{xy}},e^{\pm i \phi_{x'y'} }  \right\} = 0$ between the Dirac fermions on distinct wires.

In addition, we make sure that the ETCR is consistent with the symmetries. Under the time-reversal symmetry $T_A$, the imaginary part changes the sign, which is given as,
\begin{gather}
T_A c_{x,y,x',y'}^{a,a'}(z-z') T_A^{-1}=
\\
\nonumber
-i\pi(-1)^{\mathrm{min}\{x,x'\}+\mathrm{min}\{y,y'\}}\big[\delta_{xx'}\delta_{yy'}\delta_{aa'}\mathrm{sgn}(z'-z)
+\sigma_z\delta_{aa'}\delta_{yy'}\mathrm{sgn}(x-x')+\sigma_z\delta_{yy'}\mathrm{sgn}(a-a')-\sigma_z\mathrm{sgn}(y-y')\big]
\\
\nonumber
=-c_{x,y,x',y'}^{a,a'}(z-z')=c_{x,{y+1},x',{y+1}'}^{a,a'}(z-z').
\end{gather}
Under the glide symmetry, $G$, all terms change the sign due to $\sigma_z$ except for the sign function $sgn(z-z')$.
\begin{gather}
G c_{x,y,x',y'}^{a,a'}(z-z') G^{-1}=
\\
\nonumber
i\pi(-1)^{\mathrm{min}\{x,x'\}+\mathrm{min}\{y,y'\}}\big[\delta_{xx'}\delta_{yy'}\delta_{aa'}\mathrm{sgn}(z'-z)
-\sigma_z\delta_{aa'}\delta_{yy'}\mathrm{sgn}(x-x')-\sigma_z\delta_{yy'}\mathrm{sgn}(a-a')+\sigma_z\mathrm{sgn}(y-y')\big]
\\
\nonumber
i\pi(-1)^{\mathrm{min}\{x+1,x'+1\}+\mathrm{min}\{y,y'\}}\big[\delta_{xx'}\delta_{yy'}\delta_{aa'}\mathrm{sgn}(z-z')
+\sigma_z\delta_{aa'}\delta_{yy'}\mathrm{sgn}(x-x')+\sigma_z\delta_{yy'}\mathrm{sgn}(a-a')-\sigma_z\mathrm{sgn}(y-y')\big]
\\
\nonumber
=c_{x+1,{y},x'+1,{y}'}^{a,a'}(z'-z)
\end{gather}
The remaining translational symmetries act trivially on $c_{x,y,x',y'}^{a,a'}(z-z')$. As a result, the ETCR is consistent with the symmetry transformations defined in Eq. (\ref{CWsymmetries}).

\subsection{Explicit calculation of the Haldane nullity condition}
In this appendix, we verify the Haldane nullity condition defined in Eq. \eqref{Haldanenulity}. The nullity condition can be explicitly expanded as,
\begin{gather}
\left[\Theta^{\boldsymbol\alpha}_{x,y+1/2}(z),\Theta^{\boldsymbol\alpha'}_{x',y'+1/2}(z')\right]
=[\boldsymbol\alpha\cdot(\boldsymbol\phi^A_{x,y}(z)+\boldsymbol\phi^B_{x,y+1}(z)),
\boldsymbol\alpha'\cdot(\boldsymbol\phi^A_{x',y'}(z')+\boldsymbol\phi^B_{x',y'+1}(z'))]
\\
\nonumber
=\alpha_{a} \alpha'_{a'}[(\boldsymbol\phi^{a}_{x,y}(z)+\boldsymbol\phi^{r+a}_{x,y+1}(z)),
(\boldsymbol\phi^{a'}_{x',y'}(z')+\boldsymbol\phi^{r+a'}_{x',y'+1}(z'))]
\\
\nonumber
=\alpha_{a} \alpha'_{a'}([\boldsymbol\phi^{a}_{x,y}(z),\boldsymbol\phi^{a'}_{x',y'}(z')]
+[\boldsymbol\phi^{a}_{x,y}(z),\boldsymbol\phi^{r+a'}_{x',y'+1}(z')]
+[\boldsymbol\phi^{r+a}_{x,y+1}(z),\boldsymbol\phi^{a'}_{x',y'}(z')]
+[\boldsymbol\phi^{r+a}_{x,y+1}(z),\boldsymbol\phi^{r+a'}_{x',y'+1}(z')])
\\
\nonumber
=i\pi(-1)^{\mathrm{min}\{x,x'\}}\alpha_{a} \alpha'_{a'}\times
\\
\nonumber
((-1)^{\mathrm{min}\{y,y'+1\}}
\big[-\sigma_z\delta_{y,y'+1}-\sigma_z\mathrm{sgn}(y-(y'+1))\big]
+(-1)^{\mathrm{min}\{y+1,y'\}}\big[\sigma_z\delta_{y+1,y'}-\sigma_z\mathrm{sgn}((y+1)-y')\big])=0
\end{gather}
Therefore, the Haldane nullity condition is satisfied.

\section{Theta function of $E8$ lattice}\label{Appendix:E8}
In this section, we verify the transformation of the partition functions into the theta function of the $E_8$ lattice in Eq. \eqref{Eq:thetae8}. The elements of the $E_8$ lattice are 8 dimensional integer or half-integer vectors, which have the even sum. It can be formally written as,
\bea
\Lambda_{E_8}=\{ \vec{x}| \vec{x}\in \mathbb{Z}^8 \cup  (\mathbb{Z}+\frac{1}{2})^8 \: \textrm{and} \: \sum_{i=1}^8 x_i \equiv 0 (mod 2) \}
\eea
The theta function of the $E_8$ lattice can be decomposed into the integers and the half integer contributions. Each contribution can be re-written in terms of the theta functions of the different boundary sectors as,
\begin{gather}
\Theta_{E_8}=\sum_{x\in \Lambda_{E_8}} e^{\pi i\tau|x|^2}=\sum_{x\in \mathbb{Z}^8 \cap \sum_{i=1}^8 x_i \equiv 0 (mod 2)} e^{\pi i\tau|x|^2}+\sum_{x\in (\mathbb{Z}+\frac{1}{2})^8 \cap \sum_{i=1}^8 x_i \equiv 0 (mod 2) } e^{\pi i\tau|x|^2}
\\
\nonumber
=\sum_{x_i\in \mathbb{Z} }\frac{1}{2}(1+(-1)^{\sum_i x_i}) e^{\pi i\tau\sum_i |x_i|^2} +
\sum_{x_i\in (\mathbb{Z}+\frac{1}{2}) }\frac{1}{2}(1+(-1)^{\sum_i x_i})  e^{\pi i\tau \sum_i|x_i|^2}
\\
\nonumber
=\frac{1}{2}[\sum_{x_i\in \mathbb{Z} }e^{\pi i\tau\sum_i |x_i|^2}+\sum_{x_i\in \mathbb{Z} } (-1)^{\sum_i x_i}e^{\pi i\tau\sum_i |x_i|^2}
+\sum_{x_i\in (\mathbb{Z}+\frac{1}{2}) }e^{\pi i\tau\sum_i |x_i|^2}+\sum_{x_i\in (\mathbb{Z}+\frac{1}{2}) } (-1)^{\sum_i x_i}e^{\pi i\tau\sum_i |x_i|^2}]
\\
\nonumber
=\frac{1}{2}[\prod_{i=1}^8\sum_{x_i\in \mathbb{Z} }e^{\pi i\tau  |x_i|^2}+\prod_{i=1}^8\sum_{x_i\in \mathbb{Z} } (-1)^{ x_i}e^{\pi i\tau |x_i|^2}
+\prod_{i=1}^8\sum_{x_i\in \mathbb{Z} }e^{\pi i\tau\sum_i |x_i+\frac{1}{2}|^2}+\prod_{i=1}^8\sum_{x_i\in \mathbb{Z} } (-1)^{x_i+\frac{1}{2}}e^{\pi i\tau\sum_i |x_i+\frac{1}{2}|^2}]
\\
\nonumber
=\theta[0,0](\tau)^8+\theta[0,1/2](\tau)^8+\theta[1/2,0](\tau)^8+\theta[1/2,1/2](\tau)^8,
\end{gather}
which verifies Eq. \eqref{Eq:thetae8}.

The theta function of the $E_8$ lattice, $\Theta_{E_8}$, can be transformed into the theta function of the Minkowski embedded $E_8$ lattice. To do so, we rewrite $\Theta_{E_8}$ in the basis of the simple roots as,
\begin{gather}
\Theta_{E_8}=\sum_{x\in \Lambda_{E_8}} e^{\pi i\tau|x|^2}=\sum_{ \vec{n}\in \mathbb{Z}^8} e^{\pi i\tau|\sum_{i=1}^8 n_i \alpha_i|^2}
=\sum_{ \vec{n}\in \mathbb{Z}^8} e^{\pi i\tau(\sum_{i=1}^8 n_i \alpha_i)\cdot(\sum_{j=1}^8 n_j \alpha_j)}
=\sum_{ \vec{n}\in \mathbb{Z}^8} e^{\pi i\tau(\sum_{i,j=1}^8 n_i(K_{E8})_{ij}n_j) }
\end{gather}
where $\alpha_i$ is the simple roots of the 'conventional' $E_8$ lattice defined in Eq. \eqref{E8simpleroots0}. The Cartan matrix can be written as the scalar products of the Minkowski embedded simple roots defined in Eq. \eqref{E8simpleroots} as, $\alpha_{M,i} \cdot \eta\alpha_{M,j}=(K_{E8})_{ij}$. By using this property, we transform the theta function as,
 \begin{gather}
\Theta_{E_8}=\sum_{ \vec{n}\in \mathbb{Z}^8} e^{\pi i\tau(\sum_{i,j=1}^8 n_i(K_{E8})_{ij}n_j) }
\\
\nonumber
=\sum_{ \vec{n}\in \mathbb{Z}^8} e^{\pi i\tau(\sum_{i=1}^8 n_i \alpha_{Mi})\cdot\eta (\sum_{j=1}^8 n_j \alpha_{Mj})}
=\sum_{ \vec{x}\in E_{8,M}} e^{\pi i\tau x\cdot \eta x}\equiv \Theta_{E_{8,M}}
\end{gather}
where the last line is the definition of the theta function embedded in the Minkowski metric. Moreover, we can attach the counter-propagating Dirac wires as we constructed the gapping potential in Eq. \eqref{AAT}.
\begin{gather}
\Theta_{E_{8,M}\times U(1) \times U(1)}(\tau)=\sum_{ \vec{n}\in \mathbb{Z}^{10}} e^{\pi i\tau(\sum_{i=1}^{10} n_i \alpha_{Mi})\cdot\eta (\sum_{j=1}^{10} n_j \alpha_{Mj})}=
\sum_{ \vec{n}\in \mathbb{Z}^{10}} e^{\pi i\tau(\sum_{i,j=1}^{10} n_i(K_{E8}\oplus \sigma_z)_{ij} n_j)}
\\
\nonumber
\sum_{ \vec{n}\in \mathbb{Z}^{10}} e^{\pi i\tau(\sum_{i,j=1}^{8} n_i(K_{E8})_{ij} n_j+\sum_{i,j=9}^{10} n_i(\sigma_z)_{ij} n_j)}
=\sum_{ \vec{n}\in \mathbb{Z}^{2}} e^{\pi i\tau(\sum_{i,j=1}^{8} n_i(\sigma_z)_{ij} n_j)}\sum_{ \vec{n}\in \mathbb{Z}^{8}} e^{\pi i\tau(\sum_{i,j=1}^{8} n_i(K_{E8})_{ij} n_j)}
\\
\nonumber
=\theta[0,0](\tau)\theta[0,0](-\tau)\Theta_{E_8,M}(\tau)
\end{gather}
Therefore, attaching the counter-propagating Dirac wire is equivalent to multiplying the product of the two theta functions, $\theta[0,0](\tau)\theta[0,0](-\tau)$. 

\end{document}